\documentclass{aa}  
\usepackage{aas_macros}
\usepackage{subcaption}
\usepackage{graphicx}
\usepackage{txfonts}
\usepackage{enumitem}

\usepackage{xcolor}
\usepackage{comment}
\usepackage[version=4]{mhchem}
\definecolor{burgundy}{rgb}{0.50,0.00,0.13}
\newcommand{\oh}[1]{{#1}}

\usepackage{scalerel}
\usepackage{tikz}
\usetikzlibrary{svg.path}

\definecolor{orcidlogocol}{HTML}{A6CE39}
\tikzset{
  orcidlogo/.pic={
    \fill[orcidlogocol] svg{M256,128c0,70.7-57.3,128-128,128C57.3,256,0,198.7,0,128C0,57.3,57.3,0,128,0C198.7,0,256,57.3,256,128z};
    \fill[white] svg{M86.3,186.2H70.9V79.1h15.4v48.4V186.2z}
                 svg{M108.9,79.1h41.6c39.6,0,57,28.3,57,53.6c0,27.5-21.5,53.6-56.8,53.6h-41.8V79.1z M124.3,172.4h24.5c34.9,0,42.9-26.5,42.9-39.7c0-21.5-13.7-39.7-43.7-39.7h-23.7V172.4z}
                 svg{M88.7,56.8c0,5.5-4.5,10.1-10.1,10.1c-5.6,0-10.1-4.6-10.1-10.1c0-5.6,4.5-10.1,10.1-10.1C84.2,46.7,88.7,51.3,88.7,56.8z};
  }
}

\newcommand\orcidicon[1]{\href{https://orcid.org/#1}{\mbox{\scalerel*{
\begin{tikzpicture}[yscale=-1,transform shape]
\pic{orcidlogo};
\end{tikzpicture}
}{|}}}}

\usepackage{makecell}

\usepackage{hyperref}
\begin{document} 

\title{The atmospheres of rocky exoplanets}
\subtitle{III. Using atmospheric spectra to constrain surface rock composition}



   \author{O. Herbort 
            \inst{1} 
            \and 
            L. Sereinig 
            \inst{1} 
          }

   \institute{Institute for Astronomy (IfA), University of Vienna,
              T\"urkenschanzstrasse 17, A-1180 Vienna\\ Corresponding author \email{oliver.herbort@univie.ac.at}
             }

   \date{Received 19.11.2024; accepted 12.05.2025}

 
  \abstract
   {The crust composition of rocky exoplanets with a substantial atmosphere can not be observed directly. However, recent developments start to allow the observation and characterisation of their atmospheres.}  
   {We aim to establish a link between the observable spectroscopic atmospheric features and the mineralogical crust composition of exoplanets. This allows to constrain the surface composition just by observing transit spectra.}
   {We use a diverse set of total element abundances inspired by various rock compositions, Earth, Venus, and CI chondrite as a basis for our bottom-to-top atmospheric model. We assume thermal and chemical equilibrium between the atmosphere and the planetary surface. Based on the atmospheric models in hydrostatic and chemical equilibrium with the inclusion of element depletion due to cloud formation theoretical transit spectra are calculated.} 
   {The atmospheric type classification allows constraints on the surface mineralogy especially with respect to sulphur compounds, iron oxides and iron hydroxides, feldspars, silicates and carbon species.
   Spectral features provide the possibility to differentiate the atmospheric types and thus allow some constraints on the surface composition.}
   {}


   \maketitle
%
\section{Introduction}\label{sec:intro}
There are currently over 7000 known exoplanets\footnote{\url{https://exoplanet.eu/home/}}, with a significant percentage being terrestrial worlds. 
Revealing a large range in planetary parameters (e.g. radius, mass, density, temperature) indicating a diverse set of atmospheric, surface, and interior compositions \citep[e.g.][]{Noack2014, Leconte2015, grenfelldiversity, lichtenberg2022, Lichtenberg2025}.
Although characterising atmospheres and surfaces of rocky exoplanets is one of the driving questions of modern astrophysics (e.g. \cite{Byrne2024} , the Astro2020 Decadal Survey \citep{astro2020} , and the European Space Agency's ESA Voyage 2050 \citep{esa2050exo, janson2022occulter, rossi2022spectropolarimetry}), its direct observation is challenging at best and for many planets impossible.
On the one hand, for airless bodies, the geological composition can in principle be constrained by reflected and emitted light \citep[e.g.][]{Madden2018, AsensioRamos2021, Alei2024, Hammond2025} as well as polarimetry \citep[e.g.][]{Rossi2017, Rossi2018}.
On the other hand, for planets with a substantial atmosphere, the link between the observables in the atmosphere (absorption from molecules, effects of aerosols (clouds and hazes)) needs to be understood.
The chemical composition of the atmosphere can be indicative of the planetary interior and surface \citep[e.g.][]{schaeferBSE, Herbort20rocky1, Ortenzi2020, Timmermann2023, Baumeister2023, Seidler2024}.
Furthermore, the cloud composition present in a planetary atmosphere obstruct the view to the surface itself \citep{superearthcloud, Hellingreview}, but can in principle be used in order to constrain surface conditions \citep[e.g.][]{Loftus2019, Herbort2022}.

Exoplanets, their atmospheres, surfaces, and interiors, are expected to be much more diverse than what we know from the terrestrial planets in the solar system \citep{grenfelldiversity}.
The secondary atmospheres of rocky exoplanets is a result of outgassing over planetary time scales. 
The composition of the outgassed material itself is diverse \citep{Guimond2023} and connected to the mantle composition, which drives the outgassing \citep{Ortenzi2020, Guimond2021, Baumeister2023b, GUilmond2024}.
In particular, the redox state of the mantle can generally influence the volatile composition which is outgassed \citep{Gaillard2021, Gaillard2022}.
The outgassed material can be used to constrain the planetary interior \citep{Spaargaren2020}.
Besides the presence of a given condensate species itself, their abundance can also largely vary from planet to planet. 
Especially the amount of water present ranges from trace amounts to thick water envelopes \citep[e.g.][]{Kite2021, Kimura2022, Rogers2025}.
Besides the presence of water, further volatiles such as sulphur are being investigated \citep{Janssen2023, Lodders2024}.

For the investigation of the mineralogical and near-crust atmospheric composition, a chemical phase equilibrium between these two is often assumed
\citep[e.g.][]{Miguel2011, schaeferBSE, Kite2016, Herbort20rocky1}.
Especially for hot rocky exoplanets, the atmosphere can be composed of vaporised rock \citep[e.g.][]{vanBuchem2023, Zilinskas2023}.
Additionally to these theoretical investigations, complementary works of vaporising rocks have been done in the laboratory \citep{Thompson2021, Thompson2023}.
The atmospheric elemental composition is also effected by cloud formation \citep[e.g.][]{Ackerman2001, Mbarek2016, Herbort2022, Hellingreview}.

Recent observations with the James Webb Space Telescope (JWST) \citep{JWST} are providing first indications for detections of atmospheres on the rocky planet 55~Cnc~e \citep{hu2024secondary}.
However, observations of smaller planet such as LHS~475b \citep{99earth_inconclusive}, TRAPPIST~1b \citep{trap1b}, TRAPPIST~1c \citep{trap1c}, and GJ~1132b \citep{Xue2024} do not show conclusive evidence of the detection of an atmosphere, but rather remain consistent with airless bodies.
Future instruments on the next generation of telescopes, such as the  Extremely Large Telescope (ELT) \citep{gilmozzi2007european}, and space missions such as PLATO \citep{Rauer2025}, Ariel \citep{Tinetti2018}, the Habitable Worlds Observatory (HWO) \footnote{\url{https://science.nasa.gov/astrophysics/programs/habitable-worlds-observatory/}}, and the Large Interferometer For Exoplanets (LIFE) \citep{LIFE} will further enhance the capabilities of the detection of atmospheres of earth-sized planets.

The detection of atmospheres of rocky planets has some major complications.
Firstly, the contrast from planet to star is small and best for low-mass stars.
Secondly, the atmosphere of the planet has to be retained. 
Especially for planets around low-mass stars, this is unlikely due to the proximity to the host star and the resulting atmospheric loss \citep[e.g.][]{VanLooveren2024, VanLooveren2025}.
Thirdly, not the entirety of the atmosphere is detectable, as the atmospheres become optically thick for high pressures and at higher altitudes if clouds or hazes are present.

In order to further understand the connection of the observable parts of the atmosphere linked to the surface composition, we investigate a diverse set of total element abundances, whose resulting atmospheric compositions are covering a large diversity in atmospheric compositions \citep[see][]{woitke_coexistance}.
The basis for the atmospheric model used in this work is a bottom-to-top equilibrium chemistry model.
The near-crust atmosphere is in chemical phase equilibrium with the surface \citep{Herbort20rocky1} and throughout the atmosphere the elemental depletion due to the removal of thermally stable cloud condensates is considered \citep[see also][]{Herbort2022}.
Based on these atmospheres, we create theoretical transmission spectra to bridge potential observables to surface compositions.

At this point it needs to be noted that the assumption of chemical and phase equilibrium between the atmosphere and crust is a limiting factor for the models presented throughout this work.
Especially for the moderately low temperatures investigated here, the systems may not reach the theoretically favourable equilibrium state.
Causes for this out-of-equilibrium state may be an atmospheric composition driven by volcanic activity or photochemistry. 
The gases released from volcanic outgassing may be in equilibrium at the outgassing temperatures but the timescales for cooling down to the surrounding atmospheric temperature is governed by kinetic chemistry, which for the equilibration to low temperatures might be too long to be relevant.
Therefore, systems with atmospheres dominated by active volcanism and temperatures of $T<700\,$K might not reach the equilibrium state  \citep{Liggins2023}.
However, given enough geological inactive time, the system of atmosphere and surface composition  should evolve towards the chemical equilibrium.
However, for modest outgassing rates, the main atmospheric components can also be described in chemical phase equilibrium. 
Although Earth experiences volcanism and further effects such as biological activity, an equilibrium model similar to those presented in this paper can be constructed which only deviates in molecules with number densities of less than 5\,ppm. This especially includes \ce{CH4}, which is not stable in the equilibrium model \citep[see Appendix A in][]{Herbort2022}.

The equilibration of the condensate phase depends on the timescales of condensation directly form the gas phase as well as the rearrangement (annealing) of the condensate phase itself.
An approximation of these timescales for different rocks is also discussed in \citet{Herbort20rocky1}.
However, the temperature threshold for assuming chemical equilibrium remains unknown. 
Understanding the modelling results with this in mind, provides an insight to potential links present in understanding the atmosphere-crust and ultimately atmosphere-interior link for the diversity of rocky exoplanets \citep[see also][]{Byrne2024}.

Investigations of deviations from this equilibrium state are beyond the scope of this work and will be investigated in the future.
A further important aspect to be noted here is photochemistry, which drives the system out of chemical equilibrium to a different stable state. 
However, the compositions of this stable state are not only dependent on the atmospheric composition, but especially on the incoming stellar radiation.
This can lead to the depletion of species such as \ce{H2O}, \ce{CH4}, or \ce{NH3} with respect to their expected presence in chemical equilibrium \citep[see e.g.][]{Kasting1982, Rugheimer2015, Hu2021}.
However, the degree to which these species are depleted is highly dependent on stellar irradiation. 
As this introduces further dimensions to the investigated parameter space, we do not include this in the work at hand, as the main focus is the investigation of the changes in surface and atmospheric composition induced by differences in total elemental composition.

The nature of chemical equilibrium solvers allows the investigation of a model including many more species compared to chemical kinetics, whereas photochemical networks are due to computational constraints mostly limited to a network based on a limited number of elements, mostly limited to a subset of CHNOPS elements \citep[compare to e.g.][]{Rimmer2016, Tsai2021, Lee2024}.
As the effect of photochemistry varies strongly with the incoming stellar radiation and the atmospheric composition, their inclusion expands the potential parameter space significantly. 
Therefore, investigating the effects of photochemistry on the atmospheric composition and atmospheric types in general is beyond the scope of this work.

In Sect.~\ref{sec:methods} we provide an overview of our atmospheric and crust modelling methods, give an overview of the atmospheric classification scheme, and introduce the model-transmission spectra generation used in this work. 
In Sect.~\ref{sec:atmdiv} the diversity of the resulting model atmospheres is shown, in Sect.~\ref{sec:clouds} their clouds, and in Sect.~\ref{sec:crust} the resulting crust composition and possible links between certain crust condensates and their atmospheres. Additionally, in Sect.~\ref{sec:spec}  we discuss the resulting transmission spectra and their observability. In Sect.~\ref{sec:graphite} we describe the peculiarities of graphite as a stable condensate at low pressures. 
Section~\ref{sec:conclusion} summarises and discusses the implications of our findings.

\section{Methods}\label{sec:methods}
In this section, we briefly describe our atmospheric model (Sect.~\ref{sec:chemmethod}) , before the parameter space for the atmospheric diversity is introduced (Sect.~\ref{sec:paraspace}). 
Afterwards, we describe how the transmission spectra are created (Sect.~\ref{sec:specmethod}). 

\subsection{Atmospheric and crust modelling}
\label{sec:chemmethod}
For modelling of atmospheric and crust compositions we use the equilibrium chemistry code \textsc{GGchem} \citep{woitke2018equilibrium}. 
Based on a given set of total element abundances, a given pressure, and a given temperature, the chemical and phase equilibrium is solved.
Based on the atmosphere-crust interaction layer \citep[see also][]{Herbort20rocky1}, we build a bottom-to-top atmosphere with chemical and phase equilibrium at each layer \citep[see also][]{Herbort2022}.
The gas phase elemental composition is used as the total element abundance in the atmospheric layer above.
Thus, all thermally stable condensates (liquid and solid) are forming and deplete the effected elemental abundances in the atmospheric layers above.
At the base layer, all thermally stable solid and liquid condensates represent the crust composition. 
The different sets of total element abundances are based on a total of 18 elements (H, C, N, O, F, Na, Mg, Al, Si, P, S, Cl, K, Ca, Ti, Cr, Mn, and Fe), which can form a total of 474 gas phase species and 213 condensates, thereof 40 liquids.
Throughout this work, we use a parametrised pressure and temperature {\it(p, T)}-profile as introduced in the following and a range of different total element abundances, reflecting a variety of different surface and atmosphere compositions.

We base our atmospheric model on the polytropic atmospheric model used in \cite{Herbort2022}.
However, as we investigate transmission spectra, it is necessary to model to higher parts of the atmosphere.
In this work, we define the top of our model atmosphere as $10^{-7}$\,bar.
A purely polytropic atmosphere for a pressure change of 7 orders of magnitude is trending towards $T = 0$\,K, which is not an accurate portrayal for realistic atmospheres.
Therefore, we introduce a modified {\it(p, T)}-structure 
\begin{align}
T = T_{\mathrm{surf}}\left(\frac{p + p_{\mathrm{shift}}}{p_{\mathrm{surf}} + p_{\mathrm{shift}}}\right)^{\kappa},
\end{align}
with the introduction of a shift pressure  $p_{\mathrm{shift}}$.
Furthermore, we use a surface temperature $T_{\mathrm{surf}}$, surface pressure $p_{\mathrm{surf}}$, and the polytropic exponent $\kappa = (\gamma - 1)/\gamma$, where $\gamma$ is the polytropic index. 
We define the shift pressure as 
\begin{equation}
p_{\mathrm{shift}} = \frac{{\chi}p_{\mathrm{surf}}}{1 - {\chi}}, 
\end{equation}
where $\chi$ is given by the ratio of the surface temperature and a minimum temperature $T_{\mathrm{min}}$ as
\begin{equation}
    \chi = \left( \frac{T_{\mathrm{min}}}{T_{\mathrm{surf}}}\right)^{1/\kappa}.
\end{equation}
We note, that due to this shift, the parameters $\gamma$ and $\kappa$ are not equivalent to the polytropic parameters determined by heat capacities, but parameters in similar nature, but different values.

In reality the composition of the atmosphere and condensation of cloud condensates can strongly influence the {\it(p, T)}-structure, due to various heating and cooling effects, including the absorption of stellar radiation and the release of latent heat during condensation.
However, in order to investigate the effects caused by purely changing the base composition, we keep the {\it(p, T)}-profile consistent for different compositions.

\begin{figure}[!t]
    \centering
    \includegraphics[width=1\linewidth]{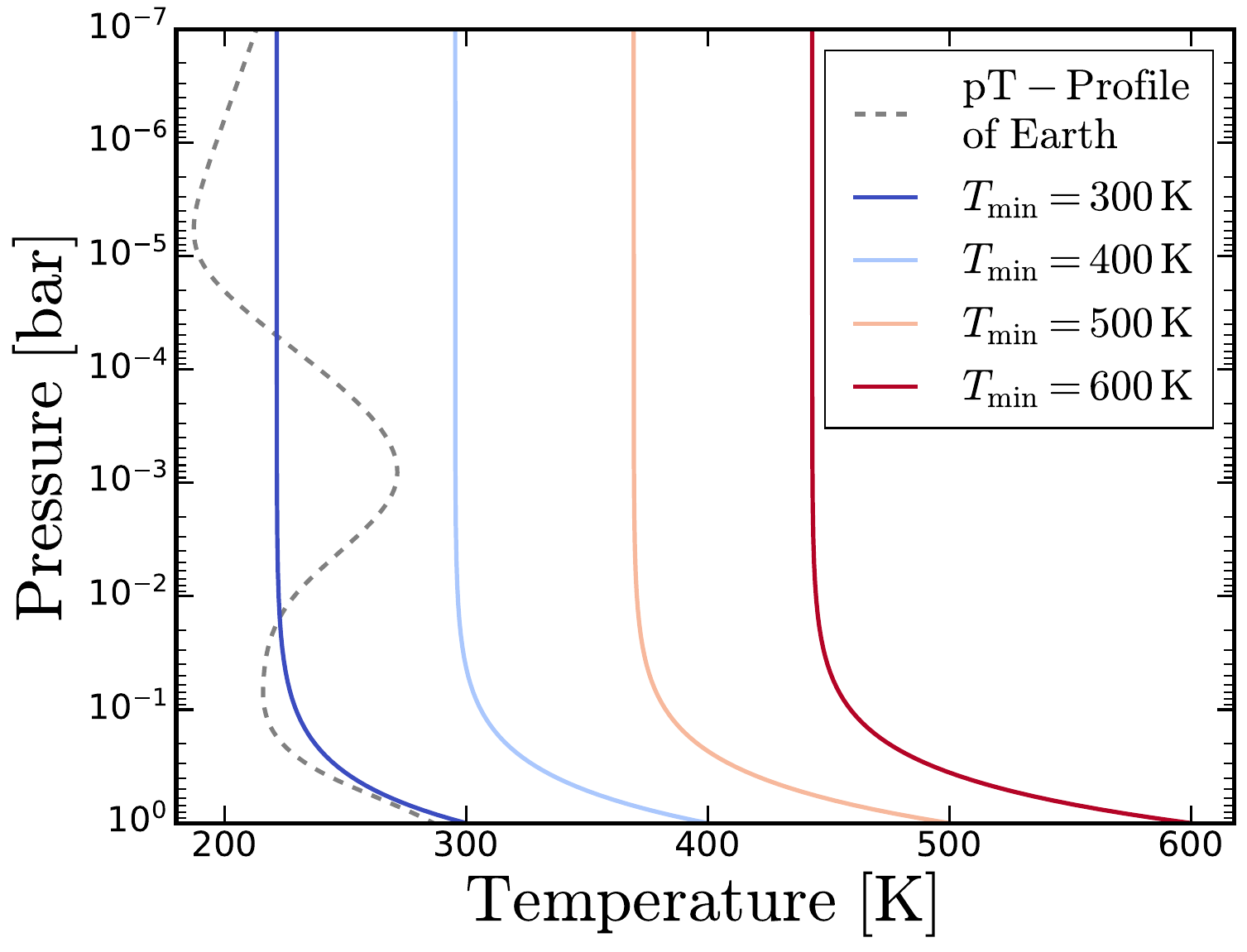}
    \caption{Comparison of our atmospheric profile model for $T_{\mathrm{surf}} = 300\,$K, $400\,$K, $500\,$K, $600\,$K, $T_{\mathrm{min}} = 0.75 \cdot T_{\mathrm{surf}}$, and $\kappa = 0.9$ in comparison to Earths atmospheric profile.}
    \label{fig:pTcomp}
\end{figure}

For the grid of {\it(p, T)}-profiles we adopt $T_{\mathrm{min}} = 0.75 \cdot T_{\mathrm{max}}$ and $\kappa = 0.9$ for all our model atmospheres.  Additionally we define four surface temperatures with $T_{\mathrm{surf}} = T_{\mathrm{max}}$ at 300\,K, 400\,K, 500\,K, and 600\,K.
Examples for the resulting atmospheric profiles can be seen in Fig.~\ref{fig:pTcomp} in comparison to Earth's atmospheric profile.

\subsection{Atmospheric diversity}
\label{sec:paraspace}
Atmospheric compositions of rocky exoplanets are expected to be more diverse than the compositions present in planets in the solar system. 
\citet{woitke_coexistance} introduce a scheme to characterise the atmospheric composition based on the abundance of the most important chemical elements (C, H, N, and O). 
This results in four distinct atmospheric types in chemical equilibrium:
\begin{align}
    \mathrm{Type~A:~} &\ce{H2O}, \ce{CH4}, \ce{NH3}, \ce{N2} \mathrm{~or~} \ce{H2}\nonumber \\
    \mathrm{Type~B:~} &\ce{H2O}, \ce{CO2}, \ce{N2}, \ce{O2}\label{eq:Types}\\
    \mathrm{Type~C:~} &\ce{H2O}, \ce{CO2}, \ce{CH4}, \ce{N2} \nonumber\\
    \mathrm{Type~D:~} &\ce{CO2}, \ce{CH4}, \ce{CO} , \ce{N2}\nonumber
\end{align}

\begin{figure}[!t]
    \centering
    \includegraphics[width=1\linewidth]{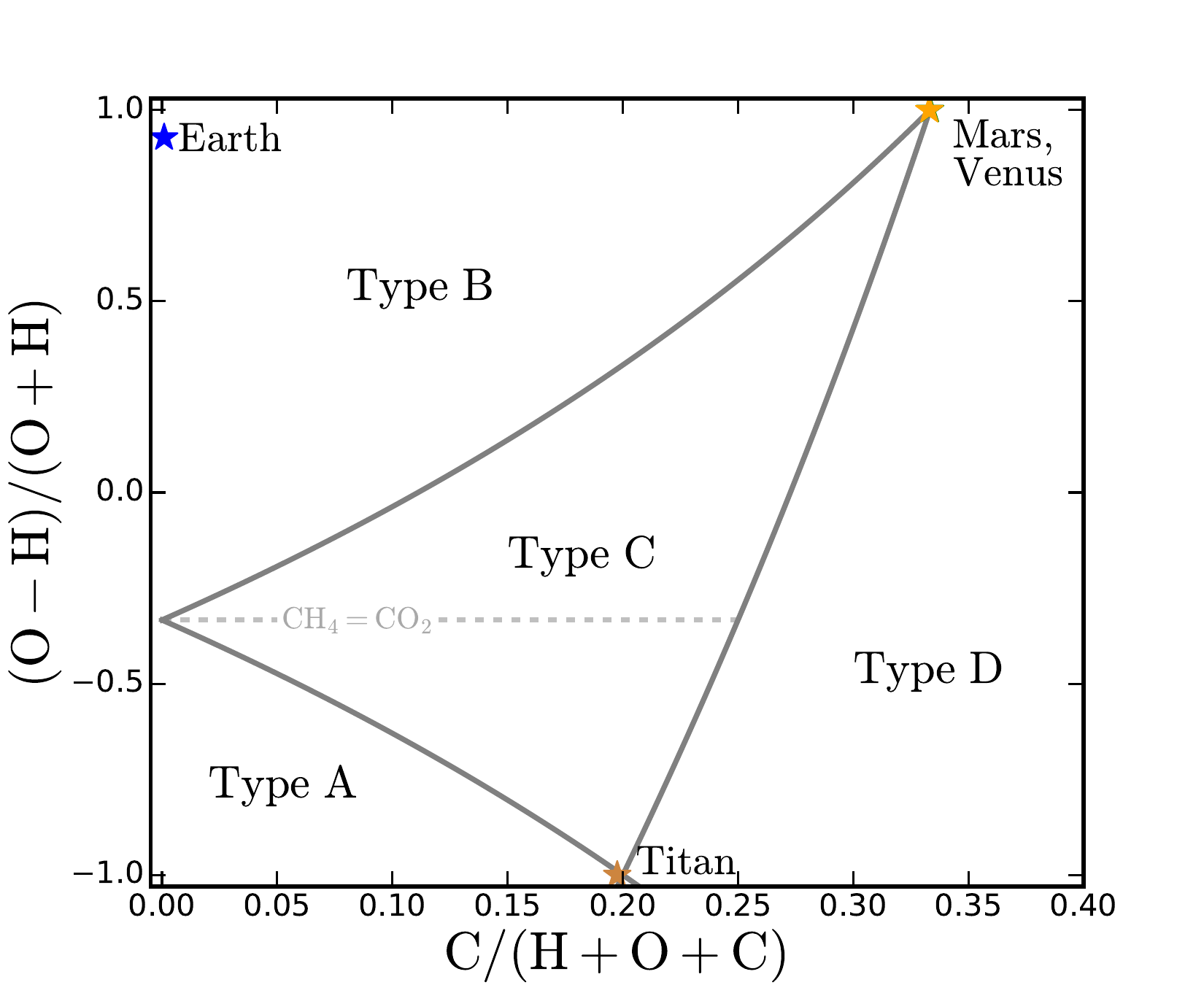}
    \caption{The parameter space for CHNO element abundances based on \cite{woitke_coexistance}. Solar system rocky bodies atmospheres are indicated with a star. 
    The different atmospheric types defined by the presence of different molecules (see Eq.~\ref{eq:Types}) are also indicated.}
    \label{fig:paraspace}
\end{figure}

In Fig.~\ref{fig:paraspace}, we show this CHO parameter space for a fixed N content. Additionally, the position of rocky solar system bodies atmospheres is shown.
The axis are given by
\begin{align}
    \frac{O -H}{O + H}, \hspace{5mm}\mathrm{and}\hspace{5mm} \frac{C}{H + O +C},
\end{align}
for the vertical axis and horizontal axis, respectively, providing a clear depiction of the composition of the different atmospheric types.

In order to cover the parameter space, we investigate different sets of total element abundances, which are given in Table~\ref{tab:abundances1} \citep[compare also to][]{Herbort20rocky1, Herbort2022}.

We use various total element abundances, based on Bulk Silicate Earth (BSE, \cite{schaeferBSE}), Mid Oceanic Ridge Basalt (MORB, \cite{arevaloMORB}), Continental Crust (CC, \cite{schaeferBSE}), CI chondrite (CI, \cite{loddersCI}) and Earth-like \citep{Herbort2022}. Furthermore, 
we investigate sets of element abundances 
which fit to Venus` atmospheric composition \citep[compare also to][]{rimmerVenus}.

In order to fill the parameter space with more diverse atmospheres we modify some of the described elemental abundances by changing the abundances of certain elements.
These changes are described in the following.

We modify the BSE abundance by adding 20\% and 30\% of H, C, and O to the initial BSE mass fractions.
These models are called BSE20HCO and BSE30HCO, respectively. Furthermore we add 8\% and 15\% water to the original BSE abundance as in \cite{Herbort20rocky1}, referred to as BSE8 and BSE15 in this work.
The resulting mass fractions can be found in Table~\ref{tab:abundances1}.

An oxygen-rich atmosphere is created based on the MORB abundance by adding 10\% mass fraction oxygen. We call the resulting model MORBo.

We explore different water contents for the Earth-like model, by reducing the H abundance by 50\% and 70\% from the initial Earth abundance to form Earth-50 and Earth-70, respectively. 
Additionally, a `dry Earth' scenario is created by reducing the H and O mass fractions of the initial Earth mass fractions such that, if all H would form water, only 10\% of the water would remain (Earthdry).

The Venus abundances are created in such a way so that they resemble gas-phase element abundances \citep{rimmerVenus}, for the venusNoSurf model (VNS). For a model which more accurately portrays the surface, we add the measurements from Vega 2 \citep{surkovVenus}. These measured references do not include the elements F, P, and Cr, so their abundance is set to 0. Because the surface elements dominate over the atmosphere, we multiply them with a factor of 1000. The model is then once again adapted to resemble Venus' atmospheric conditions and referred to as the venusSurf model (VS). We then create two more models based on this model with 2\% and 10\% additional water content (venus2 and venus10).

\subsection{Spectrum calculation with ARCiS}
\label{sec:specmethod}
For the spectra the ARtful modelling Code for exoplanet Science (\textsc{ARCiS}) code is used \citep{Min_2020}.  We input the {\it(p, T)}-structure from Sec. \ref{sec:chemmethod} and the chemical composition calculated by \textsc{GGchem}, together with planet size, mass, host star distance, and the stellar size and temperature. 
The transmission spectrum of the planetary atmosphere is then calculated based on the gas phase opacities.

However, not all molecules present in the gas phase, based on the \textsc{GGchem} calculations, have listed opacities available in {\textsc{ARCiS}}.
Additionally, due to computation time limitations, we have to constrain our compositions to the available and most abundant molecules. 
In order to compute the transmission spectra, we include all atmospheric species with peak number densities at a point in the atmosphere
of above $10^{-9}$ (1\,ppb).

The parameters for the investigated planets are resembling an Earth analogue of Earth size and Earth mass. The star-planet distance is calculated using 

\begin{equation}
    D = \sqrt{\frac{L(1 - A)}{16{\pi}{\sigma}T^4}},
\end{equation}

with $L$ being the stellar luminosity, $\sigma$ the Stefan Boltzmann constant, $A$ the planets Bond albedo, and $T$ the effective temperature. For the albedo we take the three rocky planets with an atmosphere of our solar system and take the average albedo yielding $A=0.45$, and  setting the effective temperature $T$ to $T_\mathrm{surf}$ used in the atmospheric modelling. 

The star is fixed to an M1-type star with an effective temperature of $T_\mathrm{eff} = 3660$\,K, $M = 0.42$\,M$_\odot$, and $R = 0.5$\,R$_\odot$. 
A smaller star leads to an increased transit depth, for an identical planet, making the observations easier.
Although this would make late M-dwarfs the ideal host star, it is less likely for rocky planets around late M-dwarfs to retain an atmosphere. 
This is due to the influence of the XUV on the atmosphere and the proximity of the planet to the host star \citep[e.g.][]{VanLooveren2024, VanLooveren2025}. 
Therefore we use an M1 star as a host-star for our theoretical transmission spectra.

The exclusion of clouds for the calculation of the transmission spectra provides an ideal case for the transmission spectra, as the spectral depth is not effected under the assumption of a cloud free case.
This choice is motivated because the major effect of cloud condensation, for atmospheric $(p,T)$-profiles used in this work, is limited to the lowest part of the atmosphere and may also be subject to temporal changes.
Including effects of high-altitude hazes requires the inclusion of photochemistry and is therefore subject to future work.

\section{Results}\label{sec:results}
The results are structured to first describe the atmospheric diversity of the model (Sect.~\ref{sec:atmdiv}) and cloud compositions (Sect.~\ref{sec:clouds}). Followed by a description of the link of atmospheric composition to surface condensates (Sect.~\ref{sec:crust})and with the investigation of the atmospheric spectra (Sect.~\ref{sec:spec}). 
The section concludes with an investigation of the graphite stability at low pressures (Sect.~\ref{sec:graphite}).

\subsection{Atmospheric composition}
\label{sec:atmdiv}

\begin{figure}[!t]
    \centering
    \includegraphics[width=1\linewidth]{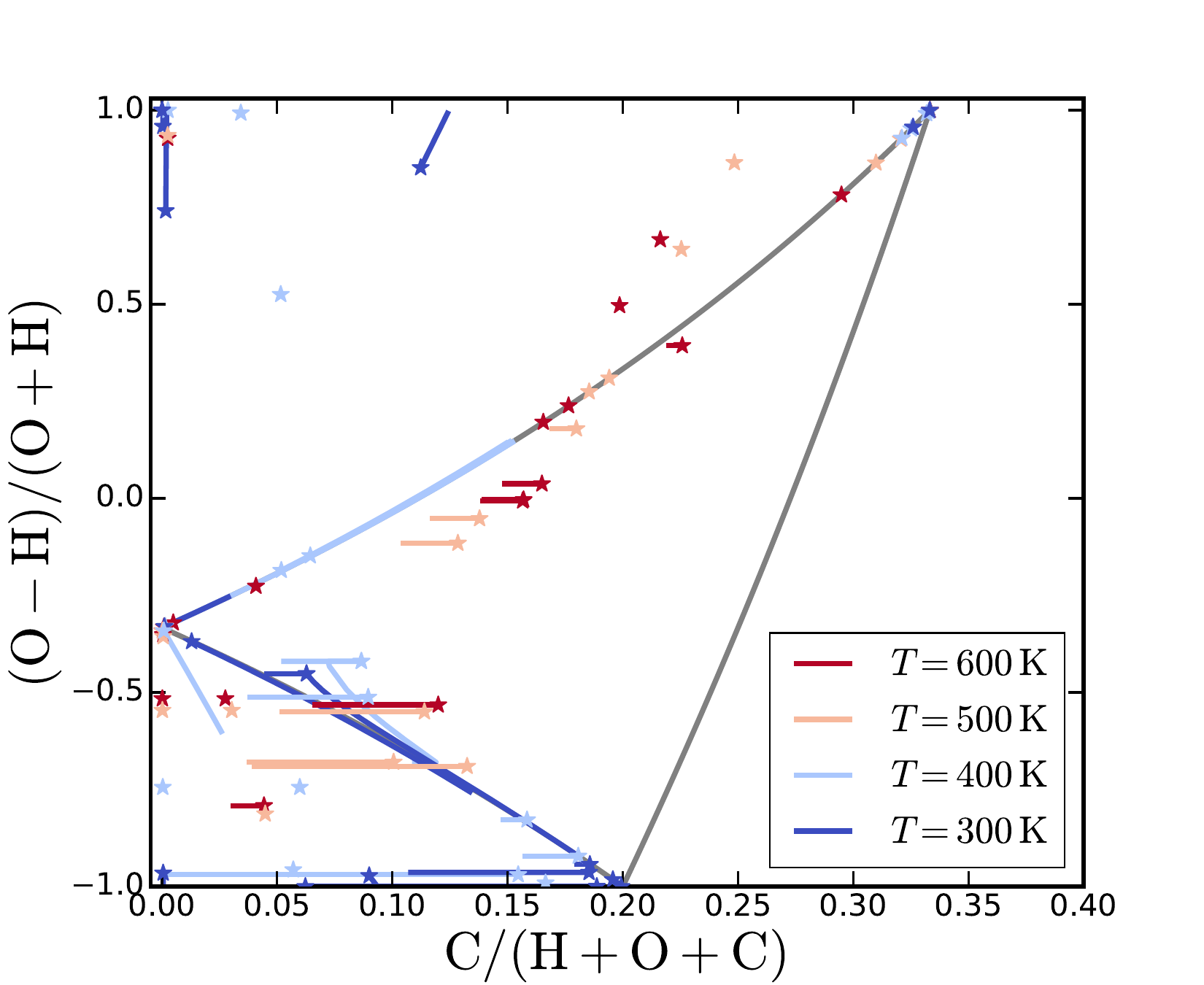}
    \caption{All atmospheres calculated in the parameter space. Some atmospheric compositions change with height due to condensation, leading to tracks. The surface level for each atmospheric track is marked with a star, surface temperature is indicated with colour.}
    \label{fig:calc_atm}
\end{figure}

The composition of an atmosphere and the corresponding atmospheric type are not fixed for one given set of total element abundances, but dependent on temperature and pressure of the system. 
Changes in these two parameters can form tracks due to the removal of thermally stable condensates from the gas phase. Figure~\ref{fig:calc_atm} shows all the calculated atmospheres, with all input abundances, at all surface temperatures in the CHO-parameter space. 
Throughout this work, we use the atmospheric types from \citet{woitke_coexistance} (see Sect.~\ref{sec:paraspace}).
All of the atmospheres are listed according to their atmospheric type in Table~\ref{tab:typestat}.
Every atmosphere is sorted by the atmospheric type of the near-crust atmosphere.

It is interesting to note that we find one type D atmosphere (venusNoSurf, 600\,K), which is - for colder temperatures - a type of composition forbidden by the supersaturation of \ce{C[s]}.
However, its high temperature of 600\,K at the surface results in \ce{CO} becoming the third most abundant molecule at low the near-crust atmosphere, surpassing \ce{H2O}.
Due to the removal of \ce{C}[s] as a thermally stable cloud condensate, this changes and the atmosphere approaches a type C atmosphere. 

\begin{table}[t]
\caption[]{\label{tab:typestat} Number of atmospheres per type and surface temperature.}
\centering
\begin{tabular}{lcccc}
\hline \hline
     & A & B & C & D\\
        \hline    
 300\,K & 8 & 5 & 8 & 0 \\
 400\,K & 7 & 5 & 9 & 0 \\
 500\,K & 6 & 5 & 10 & 0 \\
 600\,K & 5 & 6 & 9 & 1 \\
\hline
Total & 26 & 16 & 36 & 1 \\
\hline
\end{tabular}
\end{table}

In general, the atmospheric compositions found in our models are behaving like described in Sect. \ref{sec:paraspace}. We note that there are some notable extreme cases, where one molecule is by far dominating the atmospheric composition. 
In Table~\ref{tab:extremeND} we list all atmospheres for which one molecule is more abundant than 90\%. The atmospheres are sorted in columns signifying the dominating molecule species and are binned by the remaining trace gas abundance.

\begin{table*}[t]
    \caption{Atmospheres for which one molecule exceeds a mixing ratio of 90\%, sorted by the remaining cumulative mixing ratio.}
    \centering
    \resizebox{1.0\textwidth}{!}{
    \begin{tabular}{l|cccccc}
    \hline \hline
    \makecell{Trace gas\\mixing ratio} & \ce{N2} & \ce{H2O} & \ce{O2} & \ce{CH4} & \ce{CO2} & \ce{H2}   \\
    \hline
       $\leq10^{-4}$  & \makecell{BSE, 300\,K \& 400\,K \\ Earthdry, 300\,K \\ venusSurf, 300\,K \\ venus2, 300\,K \\ MORB, 300\,K \\ CC, 300\,K \\ } && MORBo, 300\,K &&&\\ \hline
       $\leq10^{-3}$  & \makecell{ BSE20HCO 300\,K \\ MORB, 400\,K} &&&&&\\ \hline
       $\leq10^{-2}$  & \makecell{venus10, 300\,K} & \makecell{venus10, 400\,K\\ venus10, 500\,K}& MORBo, 400\,K& \makecell{BSE8, 300\,K \\ archean5C, 300\,K }&&\\ \hline
       $\leq5~10^{-2}$  & \makecell{BSE, 500\,K \\ BSE30HCO, 300\,K \\  BSE15, 300\,K \\ venusSurf, 400\,K \\ venus2, 400\,K}  & \makecell{ BSE8, 500\,K\\ BSE8, 600\,K \\ BSE15 500\,K \\ BSE15 600\,K \\ venus2, 600\,K} & \makecell{Earth-50, 300\,K  \\ Earth-70, 300\,K \\  MORBo, 500\,K\\ MORBo, 600\,K }& & \makecell{Earthdry, 500\,K \\ venusSurf, 500\,K\\ venusSurf, 600\,K}& archean, 300\,K\\ \hline
       $\leq10^{-1}$  & \makecell{ Eartdry, 400\,K \\ MORB, 500\,K} & \makecell{ } & Earth-70, 400\,K && venusNoSurf, 500\,K\\ venusSurf, 600\,K &\\
       \hline 
    \end{tabular}}
    \label{tab:extremeND}
\end{table*}

\subsection{Cloud condensates}
\label{sec:clouds}

The removal of thermally stable cloud condensates is an integral part of the models presented here and allows further constraints on the surface composition. 
As the cloud model involved is purely based on the supersaturation ratio and does not include effects like nucleation  and condensation, the approach does not provide an insight to the number of cloud droplets or their size. 
Therefore we characterize the different cloud condensates by their abundance relative to the gas phase if a fixed set of element abundances is brought to lower pressures. 
This leads to four different categories of cloud condensates. 

\begin{description}
    \item[{\textsf{\textit{Main component condenses:}}}] The main species of the gas phase becomes thermally stable as a condensate. Therefore the cloud particle number density relative to the gas phase reaches values of $n_\mathrm{cond} / n_\mathrm{gas} > 0.5$ in the first condensation layer. 

    \item[{\textsf{\textit{Abundant condensates:}}}] Cloud particle number density relative to gas phase of $n_\mathrm{cond} / n_\mathrm{gas} > 10^{-9}$.

    \item[{\textsf{\textit{Trace condensates:}}}] Cloud particle number density relative to gas phase of $n_\mathrm{cond} / n_\mathrm{gas} > 10^{-12}$.

    \item[{\textsf{\textit{Numerical condensates:}}}] Within the model, there are further condensates removed from the gas phase. However, these are only present at very low number densities of less than $n_\mathrm{cond} / n_\mathrm{gas} > 10^{-12}$. Within this work, we do not further investigate such condensates, as their abundance is too low. 

\end{description}

The only species, which can condense as the main component of the atmosphere is \ce{H2O}[l], as the main gas species throughout our atmospheric models are \ce{H2O}, \ce{CO2}, \ce{CH4}, \ce{H2}, \ce{N2}, and \ce{O2} (see also Table~\ref{tab:extremeND}).
Although \ce{CO2} and \ce{CH4} have corresponding condensate phases which are included in \textsc{GGchem}, they only become thermally stable at temperatures colder than those investigated in this work, leaving \ce{H2O} as the only dominating atmospheric species to condense. This collapse of the atmosphere appears only in the models with 400\,K surface temperature, as they become cold enough for water to condense.
We note that such an atmospheric collapse is not seen in any currently known planet.
 
In the abundant cloud condensates category we find a total of four thermally stable condensates, namely \ce{H2O}[l,s], C[s], and \ce{NH4Cl}[s]. Most condensates of this category are found in atmospheres with 300\,K, 400\,K, and 500\,K respective surface temperature.

Type B atmospheres with surface temperatures of 300\,K and 400\,K exclusively condense \ce{H2O}[l,s]. We find that type A and C atmospheres can contain all of the species in the abundant condensates bin, with \ce{NH4Cl}[s] only occurring in type A and in atmospheres in the hydrogen-rich portion of type C. At 500\,K surface temperature the only cloud forming condensate, with $n_\mathrm{cond} / n_\mathrm{gas} > 10^{-12}$, is C[s], which is found in carbon-rich atmospheres  of type A and C. Type B atmospheres remain completely cloud free for $T_\mathrm{surf}=500~$K.

The models with 600\,K surface temperature give rise to a number of new condensates which are all found in trace abundances.
Overall, five different condensates form, which can be sorted into salts (\ce{NaCl}[s] and \ce{KCl}[s]), iron sulphides (\ce{FeS}[s] and \ce{FeS2}[s]), and metal oxides (\ce{Fe2O3}[s], \ce{Al2O3}[s]). 
Generally we find that C[s] is the only condensate in this temperature regime, which occurs in sufficient amounts which place it in the abundant condensate category. Furthermore C[s] is only found in atmospheres of type A, C and D. In type D C[s] is the only condensate with $n_\mathrm{cond} / n_\mathrm{gas} > 10^{-12}$. 
Type B atmospheres only condense \ce{NaCl}[s] and \ce{Al2O3}[s] in trace abundance amounts or remain cloud free. 
In type A we find the condensate species \ce{C}[s], \ce{NaCl}[s], and \ce{FeS}[s]. Hydrogen-rich type C atmospheres condensates resemble those found in type A.
C[s] and NaCl[s] are also found in oxygen-rich type C atmospheres.  While \ce{FeS}[s] is no longer found additional thermally stable cloud condensates of KCl[s], \ce{FeS2}[s], and \ce{Fe2O3}[s] emerge.
Notably \ce{Al2O3}[s] exclusively occurs in type B atmospheres.

We define an atmosphere as cloud free if there are no thermally stable condensates with a number density $n_\mathrm{cond} / n_\mathrm{gas} > 10^{-12}$. Cloud free atmospheres exist for every surface temperature with no apparent correlation with atmospheric type. We find that $\sim$33\% of atmospheres with $T_\mathrm{surf}=300~$K, $\sim$42.9\% of atmospheres with $T_\mathrm{surf}=400~$K, $\sim$61.9\% of atmospheres with $T_\mathrm{surf}=500~$K, and $\sim$23.8\% of atmospheres with $T_\mathrm{surf}=600~$K are cloud free. However, 3D effects such as temperature difference in the day and night side, which are not considered in this work, could introduce further cloud condensation.

\subsection{Atmosphere as an indicator for the crust composition}
\label{sec:crust}
One main aspect of this study is the investigation of the connection between the atmospheric and crustal composition.
Although the crustal composition for each atmosphere is diverse and a total of 83 different condensates are present throughout the investigated parameter space, the presence of some condensates can be linked to the atmospheric composition. 
The condensates with the strongest links can be separated into five distinctive groups of condensates, which are
\begin{itemize}
    \item[a)] Sulphur bearing species (\ce{FeS}[s], \ce{FeS2}[s], \ce{CaSO4}[s]),
    \item[b)] Iron oxides (evolution of redox state),
    \item[c)] Feldspar (\ce{KAlSi3O8}[s], \ce{CaAl2Si2O8}[s], \ce{NaAlSi3O8}[s]), 
    \item[d)] Silicates and Silica (\ce{Mg2SiO4}[s], \ce{MgSiO3}[s], \ce{SiO2}[s]), and
    \item[e)] Carbon compounds (\ce{C}[s], carbonates).
\end{itemize}

In the subsequent sections, these different groups are individually discussed in further detail.

\subsubsection{Sulphur bearing species}
\label{sub:sulph}
All models contain the sulphur bearing species  \ce{FeS}[s] (Iron Sulfide), \ce{FeS2}[s] (Iron Disulfide), and \ce{CaSO4}[s] (Calcium Sulfate) in one of 4 unique combinations, namely (1) \ce{FeS}[s]-only, (2) \ce{FeS2}[s]-only, (3) \ce{FeS2}[s] coexisting with \ce{CaSO4}[s], and (4) \ce{CaSO4}[s]-only.

\begin{figure}[t]
\centering
\begin{subfigure}[b]{0.5\textwidth}
   \includegraphics[width=1\linewidth]{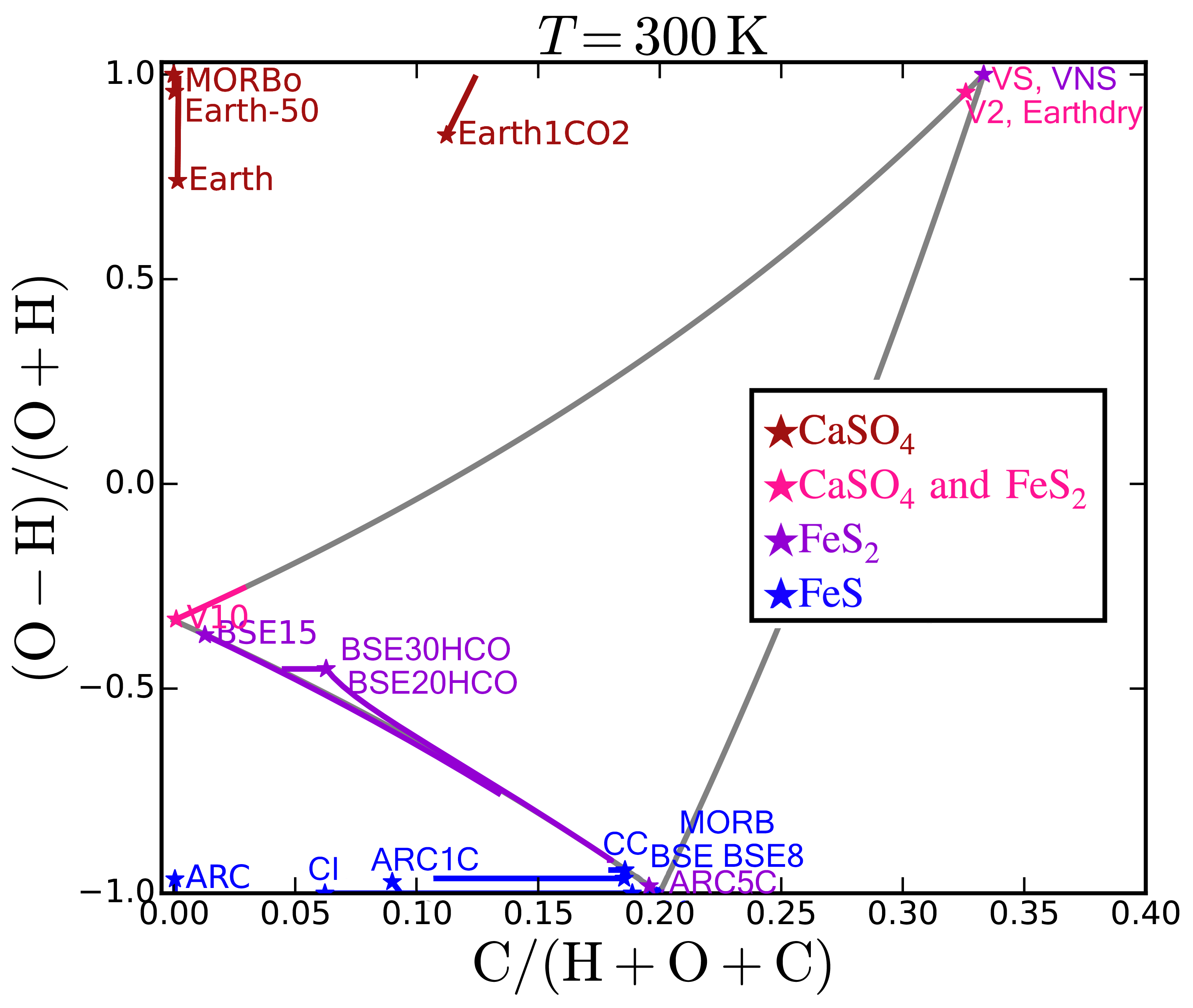}
   \label{fig:sul300} 
\end{subfigure}

\begin{subfigure}[b]{0.5\textwidth}
   \includegraphics[width=1\linewidth]{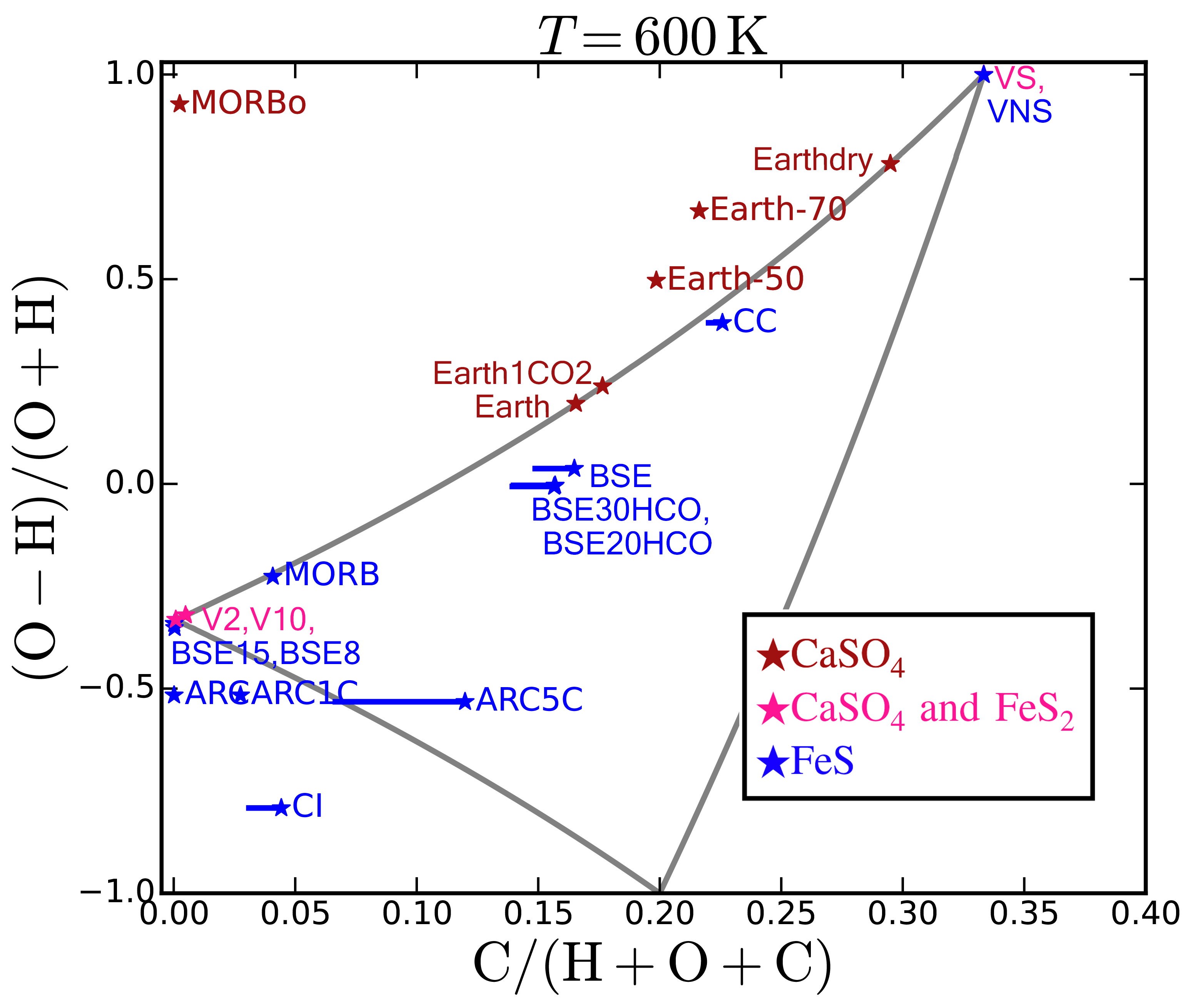}
   \label{fig:sul600}
\end{subfigure}

\caption{Sulphur bearing species in crusts at different temperatures, illustrating the transition in sulphur compounds throughout the phase space and surface temperature. The upper and lower panel are calculated for $T_\mathrm{surf}=300$\,K and $T_\mathrm{surf}=600$\,K, respectively. }
\label{fig:sulphur}
\end{figure}

At $T_\mathrm{surf}=$300\,K all but one crust of type A atmospheres have \ce{FeS}[s]-only. The one exception is the model BSE15, with an oxygen-rich atmosphere at the border from type A to type C, which has \ce{FeS2}[s]-only instead. 
Hydrogen-rich type C Atmospheres, with more \ce{CH4} than \ce{CO2}, show the presence of \ce{FeS2}[s] only. Crusts of oxygen-rich type C atmospheres with more \ce{CO2} than \ce{CH4}, have a combination of \ce{FeS2}[s] and \ce{CaSO4}[s], with the exception of the VNS model, which exclusively contains \ce{FeS2}[s] in its crust. All type B atmospheres crusts contain \ce{CaSO4}[s]-only.
The distribution of these different sulphur containing condensates is visualised in the CHO parameter space in \oh{Fig.}~\ref{fig:sul300}.

At $T_\mathrm{surf}=$400\,K the link between the sulphur compounds in the crust and the atmospheric type remains largely unchanged from the 300\,K case. 
The only difference is the transition from \ce{FeS}[s] to \ce{FeS2}[s], which no longer coincides with the type A-C border, but rather shifts towards lower hydrogen abundances, deeper into the type C atmosphere regime.

At $T_\mathrm{surf}=$500\,K the trend continues and the transition from \ce{FeS}[s] to \ce{FeS2}[s] keeps moving towards lower hydrogen abundances. All crusts of type A and type C atmospheres with \ce{CH4}$\geq$\ce{CO2} content have \ce{FeS}[s] as the sole sulphur bearing condensate in their respective crust. In crusts of type C atmospheres  where \ce{CO2}$\geq$\ce{CH4}, we find three \ce{FeS2}[s] crusts, one \ce{FeS}[s] and the remaining are of the mixed \ce{FeS2}[s] and \ce{CaSO4}[s] combination. Type B atmospheres crusts behave like before, containing exclusively \ce{CaSO4}[s].

For the highest surface temperature investigated in this work ($T_\mathrm{surf}=$600\,K), crusts with \ce{FeS2}[s] as the only sulphur bearing condensate in the crust do not occur. 
Crusts in contact with type A and C atmospheres generally only contain \ce{FeS}[s], whereas crust of type B atmospheres contain \ce{CaSO4}[s] as the exclusive sulphur bearing species.
In the transition region between type B and C atmospheres, the coexistence of \ce{FeS2}[s] with \ce{CaSO4}[s] can occur, this is also shown in Fig. \ref{fig:sul600}.

In general, the transition of the sulphur bearing species in the crust in contact with an atmospheric composition ranging from an overall reduced (hydrogen-rich) to an oxidised (oxygen-rich) atmosphere follows the transition path as
\begin{align}
    \ce{FeS}[\mathrm{s}] \longleftrightarrow \ce{FeS2}[\mathrm{s}] \longleftrightarrow \biggl(\ce{FeS2}[\mathrm{s}] + \ce{CaSO4}[\mathrm{s}]\biggr) \longleftrightarrow \ce{CaSO4}[\mathrm{s}].
\end{align}
For higher surface temperatures, the intermediate step of the sole of \ce{FeS2} is missing, with the transitions following 
\begin{align}
    \ce{FeS}[\mathrm{s}] \longleftrightarrow \biggl(\ce{FeS2}[\mathrm{s}] + \ce{CaSO4}[\mathrm{s}]\biggr) \longleftrightarrow \ce{CaSO4}[\mathrm{s}].
\end{align}
This behaviour is shown in Fig. \ref{fig:sulphur}. Throughout all temperatures, the crusts in contact with type B atmospheres contain \ce{CaSO4}[s],  with only one model showing the additional presence of \ce{FeS2}[s]. 

\subsubsection{Iron oxides and hydroxides (evolution of redox state)}
\label{sub:iron}
The crust compositions investigated in this work show the presence of various different iron containing oxides, namely \ce{FeO}[s] (Fe(II)oxide), \ce{Fe2O3}[s] (Fe(III) oxide), \ce{FeO2H}[s] (Fe(III)oxide-hydroxide), and \ce{Fe3O4}[s] (Fe(II)(III)oxide). 
The crusts have either one of these compounds by itself, or the combination of \ce{FeO2H}[s] and \ce{Fe2O3}[s], which are both Fe(III) compounds.
For some crust compositions none of these most present iron bearing compounds is thermally stable. Such a crust composition will be referred to as none (i.e. no pure iron oxide or hydroxide). 
However, there are a large number of additional Fe bearing condensates thermally stable. 
These include \ce{Fe2SiO4}[s] (fayalite),  \ce{NaFeSi2O6}[s] (aegirine), \ce{FeAl2O4}[s] (hercynite), \ce{Ca3Fe2Si3O12}[s] (andradite), \ce{FeTiO3}[s] (ilmenite), \ce{Fe}[s] (iron), \ce{FeS}[s] (ferrous sulfide), \ce{FeS2}[s] (iron disulfide), \ce{FeAl2SiO7H2}[s] (Fe-chloritoid), \ce{KFe3AlSi3O12H2}[s] (annite), \ce{Fe3Si2O9H4}[s] (greenalite), \ce{Ca2FeAl2Si3O13H}[s] (epidote), and \ce{FeCO3}[s] (siderite).

At $T_\mathrm{surf}=$300\,K and 400\,K all crusts of type A atmospheres as well as those atmospheres of type C with \ce{CH4}>\ce{CO2} have either none of the compounds or \ce{Fe3O4}[s] (Fe(II)(III)) in their crusts.
Crusts in contact with type C atmospheres containing \ce{CH4}<\ce{CO2}, contain Fe(III) compounds, with one exception (venusNoSurf, None). In crusts with type B  atmospheres \ce{Fe2O3}[s] (Fe(III)), \ce{FeO2H}[s] (Fe(III) or both are found.

At 500\,K the behaviour changes slightly for crusts in contact with type C atmospheres.
Oxygen-rich atmospheres in type C, which are close to the type B-C border, only contain \ce{Fe2O3}[s] (Fe(III)). Crusts in contact with hydrogen-rich type C or type A atmospheres, exclusively hold \ce{Fe3O4}[s] (Fe(II)(III) compound). 

At 600\,K crusts with type B and C atmospheres are the same as in crusts with cooler surface temperatures. Type A atmospheres crusts contain either none of these compounds, or \ce{FeO}[s] (Fe(II)).

Overall crusts in contact with type B atmospheres hold only Fe(III) compounds across all temperatures. Crusts of oxygen-rich type C atmospheres behave identical to crusts in contact with type B atmosphere. The rest of the phase space is more ambiguous with crusts containing either none of the mentioned compounds or a mixture of Fe(II) and Fe(III) compounds.

\subsubsection{Feldspar diversity}
\label{sub:feld}
Another group of crust condensates which show some link to the atmospheric type is the feldspar group (\ce{KAlSi3O8}[s] (orthoclase), \ce{CaAl2Si2O8}[s] (anorthite), \ce{NaAlSi3O8}[s] (albite)). 
In our model, we do not investigate solid solutions of different condensates. Although in reality such solid solutions are important, we investigate the presence and co-existence of the different endmembers.
While for all surface temperatures models without the thermal stability of any feldspar exist, the following combinations of feldspars are present for the corresponding surface temperatures:
\begin{description}
\item[{\textsf{\textit{300\,K:}}}] \ce{KAlSi3O8}[s] and \ce{NaAlSi3O8}[s] individually or together;
\item[{\textsf{\textit{400\,K:}}}] \ce{NaAlSi3O8}[s] individually or together with \ce{KAlSi3O8}[s] and/or  \ce{CaAl2Si2O8}[s];
\item[{\textsf{\textit{500\,K:}}}]  \ce{NaAlSi3O8}[s]-only, \ce{CaAl2Si2O8}[s] combined with \ce{NaAlSi3O8}[s], and all three;
\item[{\textsf{\textit{600\,K:}}}]  none, \ce{CaAl2Si2O8}[s]-only, , \ce{CaAl2Si2O8}[s] combined with \ce{NaAlSi3O8}[s], and all three.
\end{description}
We note that the only possible combination of two feldspar end-members, which we do not find to be thermally stable together in any of our models is \ce{KAlSi3O8}[s] with \ce{CaAl2Si2O8}[s].

At 300\,K surface temperature, crusts with type A atmospheres are void of any feldspar, with the exception of the MORB model, which contains \ce{KAlSi3O8}[s] combined with \ce{NaAlSi3O8}[s]. Close to the the type A-C border, in hydrogen-rich type C atmospheres crusts, we find a mixture of different feldspar content in crusts with no apparent correlations. In the remaining parameter space where \ce{CH4}<\ce{CO2}, crusts of atmospheres in type B and type C, we find \ce{KAlSi3O8}[s] combined with \ce{NaAlSi3O8}[s], with the exception of the venus2 model crust with \ce{NaAlSi3O8}[s] as the only feldspar compound.

At 400\,K surface temperature, crusts in contact with type A atmospheres are without feldspar. The only exception is the BSE model with \ce{NaAlSi3O8}[s] in the crust. In type C atmospheres where \ce{CH4}$\geq$\ce{CO2}, the respective crusts are also mostly without feldspar, except for the MORB model, conversely containing all three feldspar compounds. All other models with atmospheres of type C and B, have at least two feldspar species in their respective crusts. Most have \ce{KAlSi3O8}[s] combined with \ce{NaAlSi3O8}[s], one \ce{CaAl2Si2O8}[s] combined with \ce{NaAlSi3O8}[s] (venus2, type C atmosphere), and two oxygen rich models with all three feldspar species (venusSurf, type C atmosphere and MORBo type B atmosphere). 

For 500\,K surface temperature, type A and C have no feldspar, with one exception (BSE, type A atmosphere, \ce{NaAlSi3O8}[s]). The transition to feldspar bearing crusts happens at higher oxygen abundances than at colder temperatures, at around equal abundance of H and O.
More oxygen-rich atmospheres, both in type B and C atmospheres crusts host all three feldspar species, with one exception (venus2, type C atmosphere), with \ce{CaAl2Si2O8}[s] and \ce{NaAlSi3O8}[s].

At 600\,K the situation changes drastically, with all but two models (CI, type A atmosphere and CC type C atmosphere) crusts containing at least one feldspar species. The majority behaves as follows: crusts of atmospheres in type A have \ce{CaAl2Si2O8}[s]-only, crusts of type B have all three feldspar species, and type C is a mixture of both cases. Three (archean5C, BSE20HCO, BSE30HCO) crusts in type C atmospheres have \ce{CaAl2Si2O8}-only, one (BSE) has \ce{CaAl2Si2O8}[s] and \ce{NaAlSi3O8}[s] combined, the rest of the models has all three feldspar species combined. Those with all three species in type C are the very oxygen-richest of type C, and seem to resemble crusts of type B atmospheres.

Overall feldspar containing crusts increase with oxygen content in the respective model and with surface temperature. Type B atmospheres crusts consistently have the highest number of unique feldspar species. Which species these are, is dependent on temperature. At high temperatures (600\,K) we find feldspar species occur in almost every model, no matter what type the corresponding atmosphere is.

In the cases were a given feldspar endmember is not thermally stable, other condensate species are incorporating the K, Na, and Ca. 
For K this link between the feldspar and the replacing condensate is the strongest, with \ce{KMg3AlSi3O12H2}[s] (phlogopite) taking its place.
Additionally, \ce{KCl}[s] (potassium chloride) or \ce{K2SiO3}[s] (potassium silicate) occur. 
Similarly, instead of the thermal stability of the Na-feldspar, \ce{NaMg3AlSi3O12H2}[s] (sodaphlogopite) can be themally stable. 
However, there are also further species including \ce{Na2SiO3}[s] (sodium metasilicate), \ce{NaF}[s] (sodium fluoride), \ce{NaFeSi2O6}[s] (acmite), and \ce{NaCrSi2O6}[s] (kosmochlor) thermally stable instead of the Na forms of feldspar and phlogopite.
For the Ca-feldspar, the link to the thermal stability of another Ca-bearing condensate is less clear and is depending on the elemental composition and temperature.
For crusts in contact with H-rich atmospheres, this does especially include \ce{CaMgSi2O6}[s] (diopside).
For all others, various hydrated minerals and \ce{Ca5P3O12F}[s] (fluorapatite) are thermally stable instead of the Ca-feldspar.

\subsubsection{Silicates and Silica}
\label{sub:silicate}

The main silicates found in the crustal compositions in our models are \ce{Mg2SiO4}[s] (forsterite), \ce{MgSiO3}[s] (enstatite), and \ce{SiO2}[s] (silica).
These are found in different combinations of (1) \ce{Mg2SiO4}[s]-only, (2) \ce{Mg2SiO4}[s] plus \ce{MgSiO3}[s], (3) \ce{MgSiO3}[s]-only, (4) \ce{MgSiO3}[s] plus \ce{SiO2}[s], and (5) \ce{SiO2}[s]-only.

At low surface temperatures (300 and 400\,K) the present forms of the crust silicates contents undergo a transition from hydrogen-rich to oxygen-rich atmospheres (type A-C-B) via
\begin{align}
    \ce{Mg2SiO4}[\mathrm{s}] \longleftrightarrow \mathrm{None} \longleftrightarrow \ce{MgSiO3}[\mathrm{s}] \longleftrightarrow \ce{SiO2}[\mathrm{s}].
\end{align}

Two type A models (BSE, MORB), one type C (venusSurf), and one type B (MORBo) are outliers. The MORB, venusSurf, and MORBo models crusts contain \ce{MgSiO3}[s] plus \ce{SiO2}[s]. The BSE models crust contains \ce{Mg2SiO4}[s] with \ce{MgSiO3}[s].
At 400\,K all models still contain the same silicate combination in their crusts as in the 300\,K case.
Notably, type B atmospheres crusts always contain \ce{SiO2}[s] as the only crust silicate and once in combination with \ce{MgSiO3}[s]. Oxygen rich type C atmospheres crusts, close neighbours to type B, tend to also contain \ce{SiO2}[s] alone or in combination with \ce{MgSiO3}[s]. Two models which do not follow that trend are the venus2 model (containing only \ce{MgSiO3}[s]) and the venus10 model (containing none of the investigated silicates).

At 500\,K the transition changes compared to the lower surface temperatures cases.
The hydrogen rich part of the phase space (where atmospheres contain \ce{CH4}$\geq$\ce{CO2}), respective crusts hold \ce{Mg2SiO4}[s], with two exceptions, namely the venus10 models crust containing none of the aforementioned compounds and the BSE models crust containing \ce{Mg2SiO4}[s] plus \ce{MgSiO3}[s].
Type B and type C oxygen-rich atmospheres crusts behave the same as described in the low temperature regime.

At 600\,K the large majority of crusts contain two silicate species. There are three models with crusts that contain only one compound: CI (type A, \ce{Mg2SiO4}[s]), CC (type C, \ce{Mg2SiO4}[s]), and venusNoSurf (type D, \ce{MgSiO3}). The rest follows a transition from the hydrogen-rich to oxygen-rich part of the parameter space:
\begin{align}
    \biggl(\ce{Mg2SiO4}[\mathrm{s}] + \ce{MgSiO3}[\mathrm{s}] \biggr)\longleftrightarrow \biggl(\ce{SiO2}[\mathrm{s}] + \ce{MgSiO3}[\mathrm{s}]\biggr).
\end{align}

The transition seems to happen right at the type C-B border, with oxygen-rich type C atmospheres crusts close to the border taking on the in type B atmospheres crusts prevalent \ce{SiO2}[s] combined with \ce{MgSiO3}[s] crust silicate combination.

In summary crusts in contact type B atmospheres always have \ce{SiO2}[s], either exclusively or in combination with \ce{MgSiO3}[s] across all examined temperatures. Close to the border between type B and C, oxygen-rich type C atmospheres crusts tend to have the type B atmospheres crusts-typical silicates in their crusts. The rest of the parameter space evolves with surface temperature, from none of the listed silicates at low temperatures to containing \ce{Mg2SiO4}[s]-only and \ce{Mg2SiO4}[s] combined with \ce{MgSiO3}[s] at higher temperatures.

\subsubsection{Carbon bearing species}\label{sec:carbonsolid}

Pure carbon in the crust can only be found in type A and C atmospheres. 
The elemental carbon content of the atmosphere is then directly given by the supersaturation of C[s]. In accordance with this, no atmospheres exist in the region of CO dominated atmospheres (type D). 
However, there is one exception (VenusNoSurf, 600\,K), which shows \ce{CO}>\ce{H2O} and therefore falls into type D with higher C abundances. With the cooling along the {\it(p, T)}-profile, \ce{C}[s] is removed as a condensate and the atmosphere becomes type C.

Carbonates (\ce{CaCO3}[s] (calcium carbonate), \ce{H2CO3}[s] (carbonic acid), \ce{MgCO3}[s] (magnesium carbonate), \ce{MnCO3}[s] (manganese carbonate), \ce{FeCO3}[s] (ferrous carbonate), and \ce{NaAlCO5H2}[s] (dawsonite)): Crust carbonates predominantly occur in low type C atmospheres crusts with $T_\mathrm{surf}\leq400$\,K.

At 300\,K there are seven crusts with carbonate compounds, of which four are hydrogen-rich type C atmospheres crusts. These four models are: archean5C and venus10 with \ce{CaCO3}[s] and \ce{MnCO3}[s], BSE20HCO with \ce{MnCO3}[s] and \ce{FeCO3}[s], and BSE30HCO with \ce{MnCO3}[s], \ce{FeCO3}[s], and \ce{MgCO3}[s]. Besides those there are: one oxygen-rich type C model with the same carbon condensates as BSE30HCO, one type A atmospheres crust with \ce{CaCO3}[s], and one type B atmospheres crust with \ce{NaAlCO5H2}[s]. 

At 400\,K the situation is similar with the majority of crust carbonates occurring in low type C models. In total six crusts contain carbonates, three are hydrogen-rich type C atmospheres crusts. The three are: BSE20HCO and BSE30HCO, both containing \ce{MnCO3}[s], \ce{FeCO3}[s] and \ce{MgCO3}[s] in their respective crusts and archean5C with \ce{CaCO3}[s] in the crust. Similar to the 300\,K case, we observe one oxygen-rich type C model with the same carbon condensate as BSE20HCO and BSE30HCO, one type A atmospheres crust with \ce{CaCO3}[s], and one type B atmospheres crust with \ce{CaCO3}[s].

For higher crust temperatures crust carbonates seemingly cease to exist with only one model (CC, low type C, \ce{CaCO3}[s]) at 500\,K and one model (CC, high type C, \ce{CaCO3}[s]) at 600\,K.

\subsubsection{Further condensates}
\label{sec:restcond}

Furthermore a total of 63 additional condensates are found to be thermally stable throughout the investigated parameter space. 
These include among others hydrogen bearing condensates (phyllosilicates and liquid water), salts, and phosphor compounds.

\paragraph{Water}
The formation of phyllosilicates (hydrated minerals) is of vital importance for the presence of water as a long time thermally stable condensate.
Such minerals are often thermally favourable to liquid water \citep[e.g.][]{Herbort20rocky1}.
These can form over geological times \citep[see Mars][]{Poulet2005}, but also already during the grain formation in the protoplanetary disk \citep[e.g.][]{Thi2020}.

Our models show the presence of phyllosilicates in all atmospheric types. Naturally, due to the incorporation of OH groups, phyllosiciates are more frequent and diverse in type A and hydrogen-rich type C atmospheres.
We show the possibility that crusts in every atmospheric type can show the stability of water as a crust condensate.
We note that independent on the presence of liquid water as a condensate as part of the crust composition and the atmospheric type, water as a cloud species can be present.
Therefore simply water vapour or even clouds are not necessarily an indicator for surface water.
 
\paragraph{Salts}
Throughout the models investigated in this work, salts are an omnipresent part of the crust composition.
They include sodium chloride \ce{NaCl}[s] (sodium chloride),  \ce{KCl}[s] (potassium chloride), \ce{MgF2}[s] (magnesium flouride), \ce{CaF2}[s] (calcium flouride), \ce{NaF}[s] (sodium flouride), \ce{NH4Cl}[s] (ammonium chloride), and \ce{AlF6Na3}[s] (trisodium hexafluoroaluminate). \ce{NaCl}[s] occurs in almost every crust, with only one exception (CC, type A, \ce{KCl}[s] instead) at 300\,K surface temperature, and at 400\,K with only two exceptions (CC, type A,  \ce{KCl}[s] and venus10, type C, forms none of the compounds). For crusts with $T_\mathrm{surf}\geq500$\,K less salts are formed in contact with type type A atmospheres and hydrogen-rich type C, while crusts of oxygen-rich type C atmospheres and type B atmospheres contain \ce{NaCl}[s] in combination with \ce{MgF2}[s].

\paragraph{Phosphorus}
One element of limiting importance for the formation of life is phosphorus. 
 As also seen in \citet{Herbort2024}, all phosphorus is kept in the crust condensates in the form of hydroxy- and fluorapatite (\ce{Ca5P3O13H}[s] and \ce{Ca5P3O12F}[s]).
This is independent of the type of the connected atmosphere. \ce{Ca5P3O12F}[s] is the principle molecule and \ce{Ca5P3O13H}[s] only occurs alongside it.

\subsection{Theoretical transmission spectra}
\label{sec:spec}

\begin{figure*}[!t]
    \sidecaption
    \includegraphics[width=18cm]{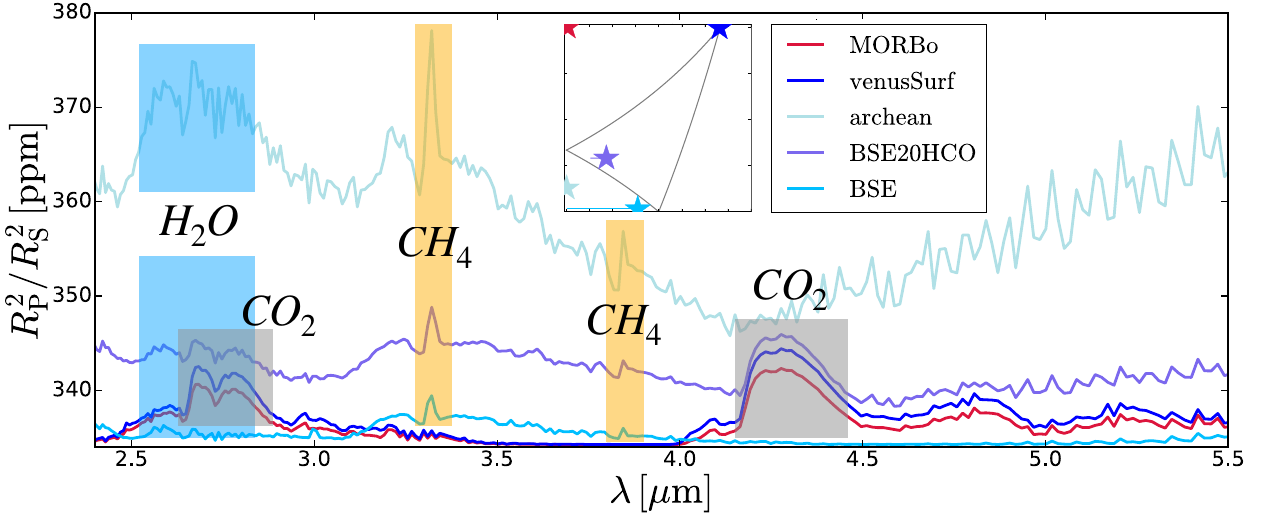}
    \caption{Comparison of transit spectra for three atmospheric types. Type A atmospheres spectra stand out with high transit depth, \ce{CH4} features at 3.3\,µm, and lack of \ce{CO2} features. Spectra originating from atmospheres of type B lack \ce{CH4} features. Spectra obtained from type C can, if oxygen-rich look like type B spectra, or, if hydrogen rich have additional \ce{CH4} features. }
    \label{fig:400Kspec}
\end{figure*}

\begin{figure*}[t]
    \sidecaption
    \includegraphics[width=18cm]{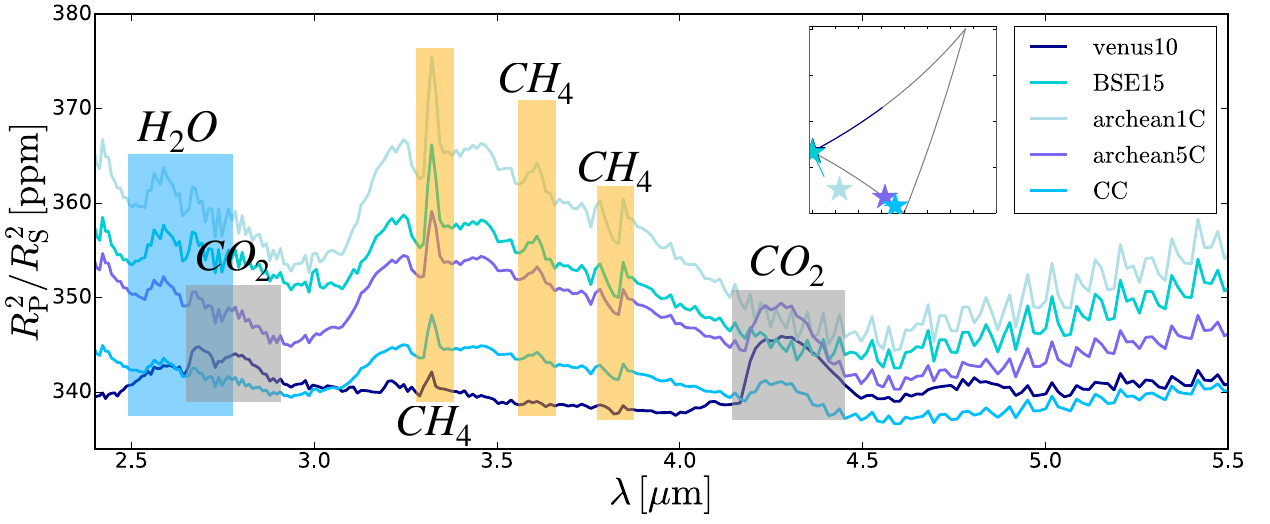}
    \caption{Comparison between spectra from type A and C atmospheres. All spectra have a sharp \ce{CH4} feature at 3.3\,µm. Spectra originating from atmospheres of type C show a defined \ce{CO2} feature at around 4.3\,µm, which can be used to differentiae the two types. However in the CC (type A) model a soft \ce{CO2} feature emerges at 4.3\,µm. }
    \label{fig:400AC}
\end{figure*}

In Section \ref{sec:crust} we have investigated the link between certain crust forming condensates and the atmospheric types defined by atmospheric composition. In this section we investigate in how far the different atmospheric types can be differentiated with transmission spectroscopy. 
Overall the main contribution to the spectra are provided by the main gas phase components of \ce{CO2}, \ce{CH4}, \ce{NH3}, and \ce{H2O}. 
Other species, especially \ce{H2}, have a significant impact on the mean molecular weight and thus on the feature depth. 
Different spectra with some marked features can be seen in Fig.~\ref{fig:400Kspec}.

In general, atmospheres of type A can be identified with a spectra by (1) their lack of strong \ce{CO2} features at $T_\mathrm{surf}\leq400$\,K, (2) \ce{CH4} features, (3) a prominent double \ce{NH3} feature at around 11\,µm, and (4) often low mean molecular weight due to higher hydrogen content resulting in inflated atmospheres and larger than usual transit depths. The fourth criterion becomes less of a distinguishing factor with higher temperatures. This is caused by the fact that atmospheres (with or without an abundance of hydrogen) show larger absorption depth with increased temperature due to the larger scale height. The largest differences in transit depth due to low mean molecular weight in type A atmospheres spectra are around 100\,ppm for the lowest $T_\mathrm{surf}$ of 300\,K. For higher $T_\mathrm{surf}$ the peak to valley difference shrinks to about 50\,ppm. For $T_\mathrm{surf}\geq500$\,K \ce{CO2} features start to appear in the spectrum, making the differentiation between type A and C atmospheres more difficult. This degeneracy can be solved by a \ce{NH3} feature at around 11\,µm, which does not appear in spectra originating from type C atmospheres.

Spectra of atmospheres in type B show signals of (1) strong \ce{CO2} features, (2) \ce{H2O} features in all spectra with $T_\mathrm{surf}\geq400$\,K and (3) lack of \ce{CH4} features. Due to their small transit depths, spectra of type B atmospheres  are the most difficult to observe. The strongest features (\ce{CO2}, \ce{H2O}) are $\sim$10\,ppm strong at 300\,K, while they roughly double for hotter $T_\mathrm{surf}$.

\begin{figure*}[t]
    \sidecaption
    \includegraphics[width=18cm]{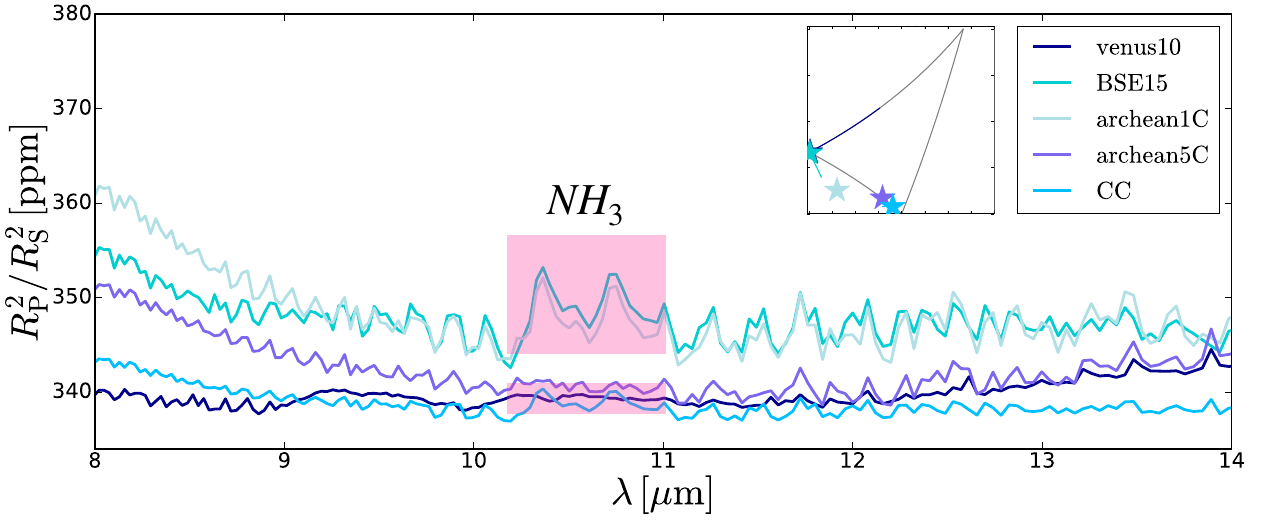}
    \caption{Comparison between spectra from type A and C atmospheres. When \ce{CO2} start to emerge in type A atmospheres at higher surface temperatures, spectra resemble those of type C atmospheres. It is however possible to use a \ce{NH3} feature in type A atmospheres spectra, to retrieve the atmospheric type from spectra alone. }
    \label{fig:400NH3}
\end{figure*}

\begin{figure*}[t]
    \sidecaption
    \includegraphics[width=18cm]{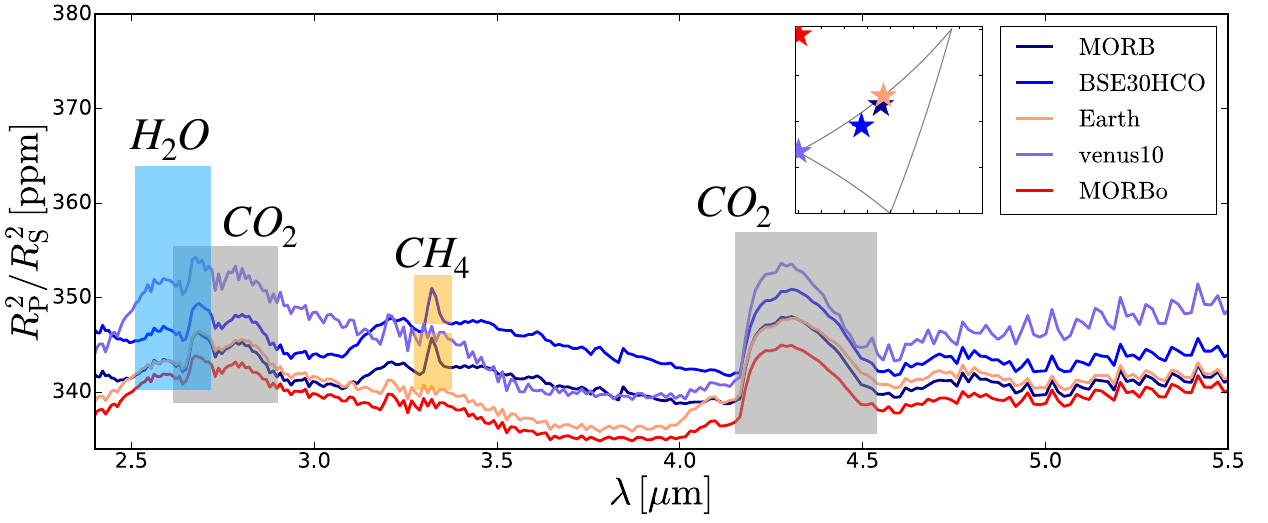}
    \caption{Displaying the sudden change in spectral features for oxygen-rich atmospheric spectra. Type B atmospheres spectra do not show any \ce{CH4} features, as do type C which are extremely oxygen-rich. Both show similar spectra, with \ce{CO2} features. In spectra from  hydrogen rich, 'deeper' type C atmospheres, strong \ce{CH4} appear.}
    \label{fig:500BC}
\end{figure*}

Type C atmospheric spectra are more ambiguous. Hydrogen-rich type C atmospheres tend to take on traits from type A atmospheres spectra, e.g. \ce{CH4} features. In that case they are still possible to tell apart from typical type A atmospheres spectra via \ce{CO2} features for cool surface temperatures below 500\,K (Fig.~\ref{fig:400AC}). As mentioned above type A atmospheres spectra can display \ce{CO2} features for higher temperatures which would make distinguishing between the two types from spectra alone impossible.  However, spectra from type C atmospheres never show \ce{NH3} features, which are exclusive to type A (Fig.~\ref{fig:400NH3}). Oxygen-rich type C atmospheres spectra behave entirely different. These spectra can be void of any \ce{CH4} features and can mimic type B atmospheres spectral features with no distinguishable qualities, despite originating from type C atmospheres. Only the very oxygen-richest type C atmospheres spectra are affected by this, as shown in Fig. \ref{fig:500BC}. When oxygen becomes less abundant \ce{CH4} features arise quickly.

The only spectrum of a type D atmosphere resembles those of type C atmospheres. 
As the type D is only found in the near-crust atmosphere and due to the removal of C as \ce{C}[s], the atmosphere transitions to an oxygen-rich type C.

\subsection{Graphite stability}
\label{sec:graphite}

\begin{figure}[t]
    \centering
    \includegraphics[width=1\linewidth]{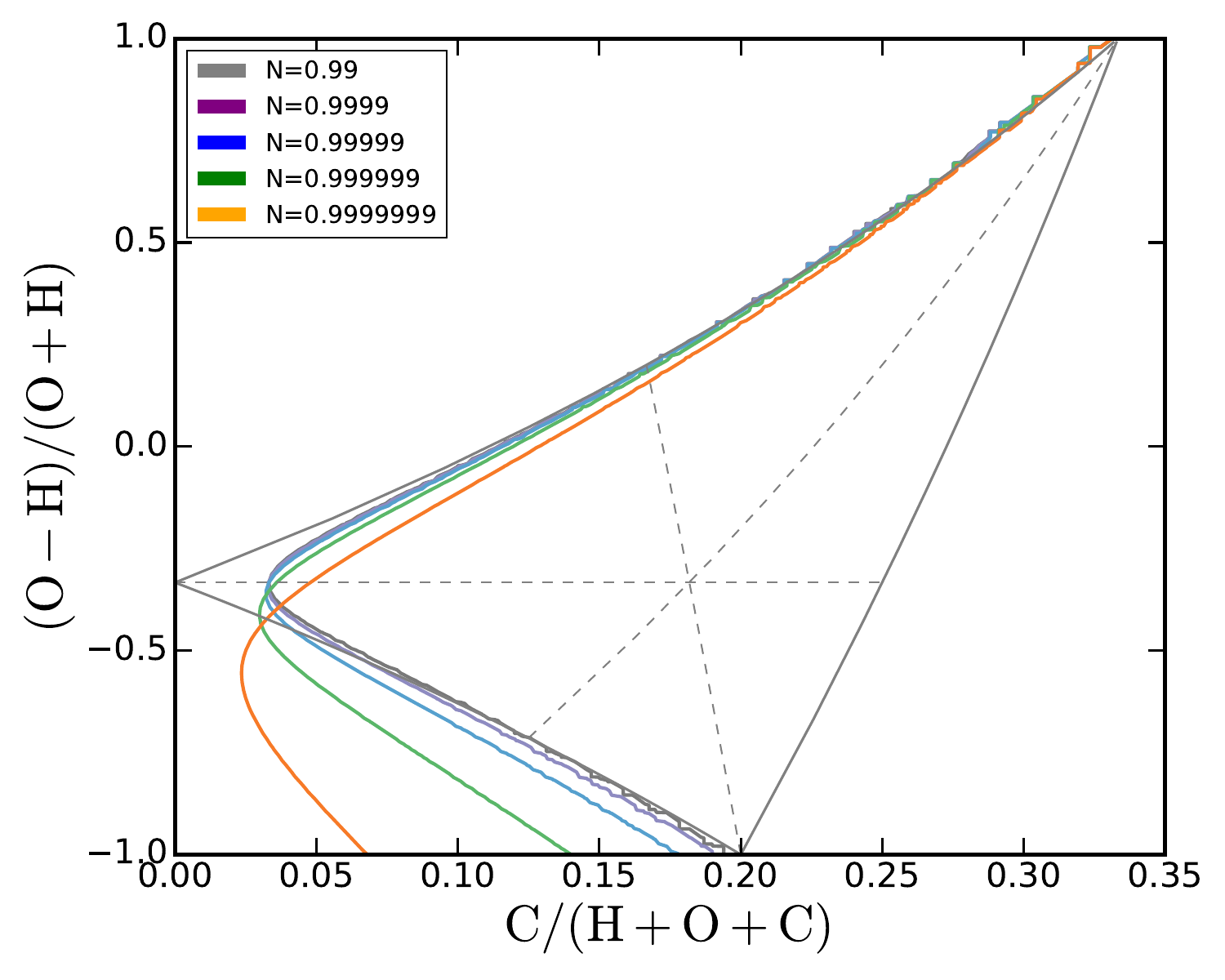}
    \caption{Influence of the nitrogen elemental abundance on the \ce{C}[s] supersaturation ratio. The different lines indicate the graphite supersaturation ratio of unity for different abundances of nitrogen. The pressure is fixed for all models to $10^{-5}$\,bar.}
    \label{fig:Csstability}
\end{figure}

We observe that the condensation of \ce{C}[s] as a thermally stable cloud condensate is not only unique because it can occur for any surface temperature, but it is also the only thermally stable condensate in the isothermal part of the atmospheric profile.
Here, the removal of purely carbon also has the effect that the atmospheric type can change from type C to type A atmospheres. 
The nature of this behaviour at low pressures and low partial pressures of carbon (i.e. high abundances of nitrogen) is visualised in Fig.~\ref{fig:Csstability}.
Here, the contour lines of the supersaturation of \ce{C}[s] of unity are indicated for different nitrogen abundances throughout the CHO-parameter space.

\section{Summary}\label{sec:conclusion}
In this paper we model planetary surfaces and atmospheres, together in a linked bottom-to-top model. 
The crust-atmosphere interaction layer is in chemical and phase equilibrium.
Throughout the atmosphere, a parcel of air follows a simple {\it(p, T)}-structure and the chemical and phase equilibrium is solved. Any thermally stable condensate throughout the atmosphere is removed as a cloud condensate and depletes the effected elements in the atmosphere above.
Our models are investigated for earth sized planets with four different surface temperatures. 
Based on these atmospheric compositions, transmission spectra are calculated with the planets orbiting an M1 type star at distances corresponding to their surface temperature.
This approach allows us to investigate the link between observables and surface compositions for a diverse set of surface and atmospheric compositions.

We explored links between atmospheric type and crust composition. Although the link is in general ambiguous and non unique, it is possible to define groups of stable crust condensates which show links between the corresponding atmospheres type and crust composition. 
Especially for crusts in contact with the oxygen-rich atmospheric type B stand out with sharply defined boundaries for respective crust compositions across all surface temperatures. Composition of crusts in contact with atmospheres of the other types (A, C, and D) show a temperature dependent behaviour, with boundaries between certain crust compositions evolving with surface temperature, at times not adhering to atmospheric type borders. 

Crust sulphur, occurring in the forms of \ce{FeS}[s], \ce{FeS2}[s], and \ce{CaSO4}[s], is tightly linked with the contacting atmosphere. We find that crusts of type B atmospheres always contain \ce{CaSO4}[s], only rarely in combination with \ce{FeS2}[s]. In contrast, no type A or D atmospheres crust contains \ce{CaSO4}[s] alone, these crusts contain \ce{FeS}[s] or \ce{FeS2}[s]. Crusts in contact with type C atmospheres can display either of the compound combinations, depending on oxygen and hydrogen content and temperature. 
Sulphur chemistry in general seems to be a promising candidate for atmosphere crust links.
\ce{H2SO4}[l,s] cloud condensates indicate high surface pressure and temperatures \citep{Herbort2022} and are incompatible with liquid water as a surface condensate \citep{Loftus2019, Jordan2025}.

Iron oxides and hydroxides show a similar behaviour, with more oxygen-rich models also containing iron in a more oxidised state. Crusts with type B atmospheres bear iron(III) compounds, while crusts in contact with type A and C atmospheres tend to contain the less oxidised iron(II) or iron(II)(III) compounds.
Showing a range of different oxidisation states expected for different planetary atmospheres \citep{Guimond2023, Nicholls2024}.
Furthermore, it should be noted that the oxidisation state of the condensates for a constant O/H can drastically differ.

For the further condensates, we find that silica (\ce{SiO2}[s]) and the silicate species \ce{Mg2SiO4}[s] and \ce{MgSiO3}[s] show links to the corresponding atmospheric type. The number of species present increases with temperature.
The number of feldspar endmembers is in general increasing with higher oxygen abundances (type B, type C with \ce{CH4}<\ce{CO2}).
Although these trends are presented in the models here, we note that further geological processes are likely to play a role. For example, the different feldspars are present in a solid solution instead of the pure endmember states, especially at or quenched at high temperatures (see e.g. \citet{Wood1978} or textbooks like \citet{klein2013earth} for further reading). 
Furthermore, active geological processes can reshape the mineral composition in the planetary interior. 
\textsc{GGchem} does not include solid solutions, instead geological models such as \textsc{Perple\_X} \citep{Myhill2021} can be used to follow up on the presence of different minerals in the planetary surface compositions. 

Pure C[s] is only found in crusts of type A and C atmospheres and can be linked to carbon clouds in the atmosphere. Carbonates predominantly form in crusts with low surface temperatures (300\,K and 400\,K) in crusts in contact with type C atmospheres.
While for type-C atmospheres, C[s] becomes thermally table in the lower parts of the atmosphere, for type A atmospheres, it is possible to have low-pressure thermal stability of C[s].
This low pressure regime coincides with an almost isothermal $(p,T)$-profile, which is unlikely to be present in the upper atmosphere due to stellar heating \citep[see also e.g.][]{VanLooveren2025}. An increasing temperature will reduce the supersaturation ratio for any condensible species.

Further crust condensates show either weak or no links with the respective atmosphere above. Phyllosilicates are more frequent at low temperatures, salts are most numerous at low surface temperatures and oxygen-rich atmospheres crusts, phosphor compounds show no correlation with atmospheric type, manganese(III)-oxide is most frequent in the oxygen-rich type B atmospheres crusts, and Ca-Ti-Al compounds show a weak link between atmosphere and surface.

Observing these atmospheres with transmission spectroscopy allows in principle to distinguish between the atmospheric types and therefore to constrain the surface minerals. Spectra of type A atmospheres can be characterised by (1) low mean molecular weight, (2) strong \ce{CH4} absorption features, (3) the existence of \ce{NH3} features, and (4) the lack of strong \ce{CO2} features at low surface temperatures. Spectra originating from type B atmospheres are defined by (1) strong \ce{CO2} features, (2) \ce{H2O} features for warm surface temperatures, and (3) lack of \ce{CH4} absorption features. The spectra of hydrogen-rich Type C atmospheres mimic those of type A atmospheres, but can be distinguished by the lack of \ce{NH3} absorption features.
While moderately oxygen-rich type C atmospheres are distinct from spectra for type B atmospheres, the spectra for the most oxygen-rich type C atmospheres seem to be indistinguishable from those for carbon-rich type B atmospheres.

Throughout this work, we have investigated atmospheric models, which include the effect of element depletion due to cloud condensation.
The effect of clouds on the spectra itself has been omitted, as this is a more diverse influence based, among others, on the cloud condensate density, composition, and size distribution. 
Furthermore, planets are inherently three-dimensional objects, which causes differences in the morning and evening terminator which are probed simultaneously during transmission spectroscopy. 
This effect is stronger for more pronounced day night side differences (see e.g. \citet{Helling2023} for gas giants or \citet{Nguyen2024} magma ocean planets).
Therefore, the transmission spectra presented here must be seen as an theoretical idealised scenario.

While the precision needed to distinguish the different spectra presented in this is challenging for instruments on JWST, future missions such as HWO and LIFE might be able to investigate these differences.

\begin{acknowledgements}
The authors thank R.J. Spaargaren for his valuable discussions on (exoplanet) mineralogy. 
\end{acknowledgements}

\bibliography{library.bib}
\onecolumn
\appendix
\section{Supplementary material}

\begin{table*}[!b]
\caption[]{\label{tab:abundances1}Elemental mass fractions of the sets of element abundances used in this work.}
\begin{center}
\resizebox{1.0\textwidth}{!}{
\begin{tabular}{l|ccccccccccccc}
\hline \hline
        & BSE & BSE20HCO & BSE30HCO & BSE8    & BSE15   & Earth   & Earth-50 & Earth-70 & Earthdry & Earth1CO2 &\\
        & 1   & $1\ast$  & $1\S$  & $1\dagger$ & 1\textdaggerdbl & 2       & $2\alpha$ & $2\beta$ & $2\gamma$ & $2\nabla$ &\\
\hline
H       & 0.006   & 0.6995  & 1.0463  & 0.8303  & 1.4574  & 0.3341 & 0.1671  & 0.1002  & 0.0441  & 0.3308 &\\
C       & 0.006   & 8.2829  & 12.4208 & 0.006   & 0.005   & 1.957  & 1.9603  & 1.9766  & 2.0157  & 2.2103 &\\
N       & 8.8E-05 & 0.0001  & 0.0001  & 8.2E-05 & 7.7E-5  & 0.0253 & 0.0254  & 0.0256  & 0.0261  & 0.0251 &\\
O       & 44.42   & 46.644  & 47.7055 & 47.8145 & 50.3196 & 50.1   & 50.1840 & 50.602  & 49.203  & 50.326 &\\
F       & 0.002   & 0.0016  & 0.0014  & 0.002   & 0.002   & 0.0482 & 0.0482  & 0.0486  & 0.0496  & 0.0477 &\\
Na      & 0.29    & 0.2325  & 0.2035  & 0.27    & 0.25    & 2.145  & 2.1486  & 2.1665  & 2.2094  & 2.1236 &\\
Mg      & 22.01   & 17.6480 & 15.442  & 20.42   & 19.18   & 1.999  & 2.0024  & 2.0190  & 2.059   & 1.979 &\\
Al      & 2.12    & 1.6999  & 1.4874  & 1.97    & 1.85    & 7.234  & 7.2461  & 7.3065  & 7.451   & 7.1617 &\\
Si      & 21.61   & 17.3273 & 15.1614 & 20.05   & 18.83   & 26.17  & 26.2139 & 26.4323 & 26.9551 & 25.9083 &\\
P       & 0.008   & 0.0064  & 0.0056  & 0.007   & 0.007   & 0.0691 & 0.0692  & 0.0698  & 0.0711  & 0.0684 &\\
S       & 0.027   & 0.0216  & 0.0189  & 0.025   & 0.024   & 0.0636 & 0.0637  & 0.0642  & 0.0655  & 0.063 &\\
Cl      & 0.004   & 0.0032  & 0.0028  & 0.004   & 0.003   & 0.0427 & 0.0428  & 0.0431  & 0.044   & 0.0423 &\\
K       & 0.02    & 0.0160  & 0.014   & 0.02    & 0.02    & 1.945  & 1.9483  & 1.9645  & 2.0034  & 1.9256 &\\
Ca      & 2.46    & 1.9725  & 1.7259  & 2.28    & 2.14    & 3.499  & 3.5049  & 3.5341  & 3.604   & 3.464 &\\
Ti      & 0.12    & 0.0962  & 0.0842  & 0.11    & 0.1     & 0.3644 & 0.3650  & 0.3681  & 0.3753  & 0.3608 &\\
Cr      & 0.29    & 0.2325  & 0.2035  & 0.27    & 0.25    & 0.0118 & 0.0118  & 0.0119  & 0.0122  & 0.0117 &\\
Mn      & 0.11    & 0.0882  & 0.0772  & 0.1     & 0.1     & 0.0654 & 0.0655  & 0.0661  & 0.0674  & 0.0648 &\\
Fe      & 6.27    & 5.0274  & 4.399   & 5.82    & 5.46    & 3.926  & 3.9326  & 3.9654  & 4.0438  & 3.8867 &\\
\hline
Sum     & 99.7731 & 100     & 99.9995 & 100     & 100     & 99.9996& 99.9998 & 100.7647& 100.2998& 99.9999 &\\
\hline
%
\hline
        & venusNoSurf & venusSurf & venus2 & venus10 & MORB & MORBo & archean & archean1C & archean5C & CC & CI \\
        & VNS & VS & V2 & V10 &  &  &  &  &  &  &  \\
        & 3           & 3\&4         & $3\&4\dagger$     &   $3\&4$\textdaggerdbl      & 5  & $3\delta$  & 2       & $2\epsilon$         & $2\zeta$ & 6 & 7 \\
\hline
    H  & 0.0005  & 1.203E-7 & 0.2222   & 1.1853   & 0.023   & 0.021 & 2.309   & 2.2859    & 2.1936   & 0.043 & 1.93 \\
    C  & 7.52    & 0.0201   & 0.0197   & 0.0181   & 0.019   & 0.017 & 0.0052  & 1.0051    & 5.0049   & 1.2   & 1.2\\
    N  & 1.41    & 0.0017   & 0.0017   & 0.0015   & 5.5e-05 & 5E-5  & 7.6E-05 & 7.524E-05 & 7.22E-05 & 0.006 & 0.288\\
    O  & 50.73   & 46.9339  & 47.7714  & 51.6478  & 44.5    & 50.1  & 49.88   & 49.381    & 47.386   & 39.4  & 40.1 \\
    F  & 0       & 0        & 0        &  0       & 0.017   & 0.015 & 0.0017  & 0.0017    & 0.0016   & 0.051 & 0.0057\\
    Na & 1.527   & 1.19974  & 1.9575   & 1.7977   & 2.012   & 1.829 & 0.251   & 0.2485    & 0.2385   & 2.27  & 0.488\\
    Mg & 6.671   & 8.764    & 8.5887   & 7.8876   & 4.735   & 4.305 & 19.01   & 18.820    & 18.060   & 17.3  & 17.6\\
    Al & 7.375   & 9.7022   & 9.5082   & 8.732    & 8.199   & 7.454 & 1.831   & 1.8127    & 1.7395   & 0.26  & 0.26\\
    Si & 17.78   & 23.3819  & 22.9142  & 21.0437  & 23.62   & 21.47 & 18.67   & 18.483    & 17.737   & 17.3  & 17.6\\
    P  & 0       & 0        & 0        & 0        & 0.057   & 0.03  & 0.0069  & 0.0068    & 0.0066   & 0.21  & 0.22\\
    S  & 1.1756  & 1.5525   & 1.5214   & 1.3972   & 0.11    & 0.1   & 0.023   & 0.0228    & 0.0219   & 3.2   & 3.2\\
    Cl & 5.87E-6 & 7.716E-9 & 7.562E-9 & 6.945E-9 & 0.014   & 0.013 & 0.0035  & 0.0035    & 0.0033   & 0.045 & 0.0682\\
    K  & 0.0493  & 0.0655   & 0.0642   & 0.059    & 0.152   & 0.138 & 0.017   & 0.0168    & 0.0162   & 2.06  & 0.0532\\
    Ca & 3.147   & 4.1335   & 4.0509   & 3.7202   & 8.239   & 7.49  & 2.125   & 2.1038    & 2.0188   & 6.6   & 6.7\\
    Ti & 0.0587  & 0.0767   & 0.0752   & 0.0069   & 0.851   & 0.774 & 0.104   & 0.1030    & 0.0988   & 0.385 & 0.0441\\
    Cr & 0       & 0        & 0        & 0        & 0.033   & 0.03  & 0.251   & 0.2485    & 0.2385   & 0.012 & 0.259\\
    Mn & 0.0446  & 0.0607   & 0.0595   & 0.0546   & 0.132   & 0.12  & 0.095   & 0.0941    & 0.0903   & 0.069 & 0.189\\
    Fe & 2.513   & 3.3099   & 3.2437   & 2.9789   & 7.278   & 6.616 & 5.416   & 5.3618    & 5.1452   & 9.6   & 9.7764\\ \hline

Sum & 100    & 100          & 99.9985  & 99.6302 & 99.9911  & 100.5464&99.9994& 99.9993   & 99.9994  & 100 & 100\\
\hline
\end{tabular}
}
\end{center}
$\ast$: Addition of equal parts H, C, and O to a total of 20\% mass fraction.\\
$\S$: Addition of equal parts H, C, and O to a total of 30\% mass fraction.\\
$\dagger$: Addition of 8\% mass fraction water.\quad
\textdaggerdbl: Addition of 15\% mass fraction water.\quad
$\alpha$: Subtraction of 50\% of the total available water.\\ 
$\beta$: Subtraction of 70\% of the total available water.\quad
$\gamma$: Subtraction of all the total available water.\\
$\nabla$: Addition of 1\% mass fraction $\mathrm{CO}_2$.\quad
$\dagger$: Addition of 2\% mass fraction water.\quad
\textdaggerdbl: Addition of 10\% mass fraction water.\\
$\delta$: Addition of 10\% mass fraction oxygen.\quad
$\epsilon$: Addition of 1\% mass fraction carbon.\quad
$\zeta$: Addition of 5\% mass fraction carbon.\\
{\textbf{References. }(1)~\citet{schaeferBSE}; (2) \citet{Herbort2022}; (3)~\citet{rimmerVenus}; (4) \citet{surkovVenus}; (5) \citet{arevaloMORB}; (6)~\citet{schaeferBSE}; (7)~\citet{loddersCI}.
}
\end{table*}

\begin{table*}[t]
\caption[]{\label{tab:condensates} List of all condensates in this work with their name, chemical and sum formula, and whether they are present as cloud and/or crust condensates.}
\centering
\begin{tabular}{ccccc}
\hline \hline
Name              &Chemical Formula   &Sum Formula&Cloud&Crust\\ \hline
Water             &\ce{H2O}[l,s]      &\ce{H2O}[l,s]       & x & x  \\
Graphite          &\ce{C}[s]          &\ce{C}[s]      & x & x  \\
Ammonium chloride &\ce{NH4Cl}[s]      &\ce{NH4Cl}[s]  & x & x  \\
Iron sulfide      &\ce{Fe(II)S}[s]    &\ce{FeS}[s]    & x & x  \\
Iron disulfide    &\ce{Fe(II)S2}[s]   &\ce{FeS2}[s]   & x & x  \\
Calcium Sulfate   &\ce{CaSO4}[s]      &\ce{CaSO4}[s]  &   & x  \\
Fe(II)oxide       &\ce{FeO}[s]        &\ce{FeO}[s]    &   & x \\
Fe(III) oxide     &\ce{Fe2O3}[s]      &\ce{Fe2O3}[s]  & x & x \\
Fe(III)oxide-hydroxide&\ce{FeO(OH)}[s]  &\ce{FeO2H}[s]  &   & x \\
Fe(II)(III)oxide  &\ce{Fe3O4}[s]      &\ce{Fe3O4}[s]  &   & x \\
Fayalite          &\ce{Fe2SiO4}[s]    &\ce{Fe2SiO4}[s]&   & x \\
Aegirine          &\ce{NaFeSi2O6}[s]  & \ce{NaFeSi2O6}[s]& &x \\
Hercynite         &\ce{FeAl2O4}[s]    &\ce{FeAl2O4}[s]  & & x \\
Andradite         &\ce{Ca3Fe2(SiO4)3}[s]&\ce{Ca3Fe2Si3O12}[s]  & & x \\
Ilmenite          &\ce{FeTiO3}[s]     &\ce{FeTiO3}[s]  & & x \\
Iron              &\ce{Fe}[s]         &\ce{Fe}[s]  & & x \\
Fe-Chloritoid     &\ce{Fe2Al4Si2O10(OH)4}[s]&\ce{FeAl2SiO7H2}[s]  & & x \\
Annite            &\ce{KFe(II)3AlSi3O10(OH)2}[s]&\ce{KFe3AlSi3O12H2}[s] & & x \\
Greenalite        &\ce{Fe(II)3Si2O5(OH)4}[s]   &\ce{Fe3Si2O9H4}[s]  & & x \\ 
Epidote           &\ce{Ca2Al2Fe(III)(SiO4)(Si2O7)O(OH)}[s]&\ce{Ca2FeAl2Si3O13H}[s]&& x \\
Siderite          &\ce{FeCO3}[s]   &\ce{FeCO3}[s]  & & x \\
Orthoclase        &\ce{KAlSi3O8}[s]   &\ce{KAlSi3O8}[s]  & & x \\
Anorthite         &\ce{CaAl2Si2O8}[s]   &\ce{CaAl2Si2O8}[s]  & & x \\
Albite            &\ce{NaAlSi3O8}[s]   &\ce{NaAlSi3O8}[s]  & & x \\
Fosterite         &\ce{Mg2SiO4}[s]   &\ce{Mg2SiO4}[s]  & & x \\
Enstatite       &\ce{MgSiO3}[s]   &\ce{MgSiO3}[s]  & & x \\
Silica          &\ce{SiO2}[s]     &\ce{SiO2}[s]  & & x \\
Calcium Carbonate &\ce{CaCO3}[s]   &\ce{CaCO3}[s]  & & x \\
Carbonic Acid     &\ce{H2CO3}[s]   &\ce{H2CO3}[s]  & & x \\
Magnesium Carbonate&\ce{MgCO3}[s]   &\ce{MgCO3}[s]  & & x \\
Manganese Carbonate&\ce{MnCO3}[s]   &\ce{MnCO3}[s]  & & x \\
Ferrous Carbonate &\ce{FeCO3}[s]   &\ce{FeCO3}[s]  & & x \\
Dawsonite         &\ce{NaAlCO3(OH)2}[s]   &\ce{NaAlCO5H2}[s]  & & x \\
Sodium Chloride   &\ce{NaCl}[s]   &\ce{NaCl}[s]  & x & x \\
Potassium Chloride&\ce{KCl}[s]   &\ce{KCl }[s]  &x& x \\
Magnesium Flouride&\ce{MgF2}[s]   &\ce{MgF2}[s]  & & x \\
Calcium Flouride  &\ce{CaF2}[s]   &\ce{CaF2}[s]  & & x \\
Sodium Flouride   &\ce{NaF}[s]   &\ce{NaF}[s]  & & x \\
Trisodium Hexafluoroaluminate&\ce{AlF6Na3}[s]  &\ce{AlF6Na3}[s]& & x \\
Hydroxylapatite&\ce{Ca5(PO4)3OH}[s]   &\ce{Ca5P3O13H}[s]  & & x \\
Fluorapatite&\ce{Ca5(PO4)3F}[s]   &\ce{Ca5P3O12F}[s]  & & x \\
Corundum          &\ce{Al2O3}[s]      &\ce{Al2O3}[s]  & x & x \\\hline 
\end{tabular}
\end{table*}
\end{document}